\documentclass[useAMS,usenatbib]{mnras}
\usepackage{amsmath}
\usepackage{graphicx}
\usepackage{txfonts}
\usepackage{natbib}
\usepackage{bm}
\usepackage{changes}
\usepackage{multirow}
\usepackage{caption}
\usepackage{nicematrix}
\usepackage{enumitem}
\usepackage{tabularx,booktabs}

\def\apj{ApJ}
\def\apjl{ApJL}

\def\aap{A\&A}

\def\beq#1{\begin{equation}\label{#1}}
\def\eeq{\end{equation}}
\def\beqa#1{\begin{eqnarray}\label{#1}}
\def\eeqa{\end{eqnarray}}

\def\Eq#1{Eq.~(\ref{#1})} 
\def\eqn#1{~(\ref{#1})}
\def\myfrac#1#2{\left(\frac{#1}{#2}\right)}

\def\comment#1{\relax}

\title[Orbital Period Changes in WR+OB Binaries]{Spectroscopic Searches for Evolutionary Orbital Period Changes in WR+OB Binaries: The case of CQ Cep and CX Cep}
\author[A. Cherepashchuk et al.] {I. Shaposhnikov$^{1,2}$
\thanks{E-mail: iv.shaposhnikov@gmail.com, cherepashchuk@gmail.com, kpostnov@gmail.com},
A. Cherepashchuk$^1$, A. Dodin$^1$,  K. Postnov$^{1,3}$\\
$^{1}$Moscow State University, Sternberg Astronomical Institute, Universitetskij pr. 13, 119234 Moscow, Russia\\
$^{2}$Moscow State University, Faculty of Physics, Leninskiye Gory 1-2, 119991 Moscow, Russia\\
$^{3}$Kazan Federal University, Kremlyovskaya 18, 420008 Kazan, Russia\\
}	

\begin{document}

\maketitle
\date{Received ... Accepted ...}
\pagerange{\pageref{firstpage}-\pageref{lastpage}} \pubyear{2023}

\maketitle

\label{firstpage}

\begin{abstract}
We present the results of spectroscopic observations of two eclipsing WR+OB-type systems - CQ Cep and CX Cep, performed in 2020-2023 with a low-resolution slit spectrograph TDS ($\lambda\lambda= 3660-7410$\AA, $R = 1300-2500$) on 2.5-m telescope
of the SAI MSU Caucasian Mountain Observatory.
For CQ Cep, the radial velocity curves of a WN6 star are constructed, the problem of visibility of spectroscopic traces of an OB star is discussed and 
the components' mass ratio $q\sim 0.6$ is estimated. For CX Cep, the radial velocity curves are constructed for both the WN5 and O5 components
enabling their masses and circular orbit elements to be refined. The comparison of the radial velocity curves of these systems obtained in different epochs allowed us to derive the orbital period change rate $\dot{P}$ by the spectroscopic method, which is found to be in good agreement with estimates obtained by comparing the moments of primary eclipse minima: $\dot{P} = -0.0151\pm0.0013$ s yr$^{-1}$ for CQ Cep and $\dot{P} = 0.054\pm0.009$ s yr$^{-1}$ for CX Cep. The prospects of applicability of the spectroscopic dynamical method for studying the orbital evolution of 
Galactic WR+OB binaries and related objects are considered.
We also discuss the effect of finite sizes of stars with stellar wind mass loss in close binary systems on 
their orbital evolution.
\end{abstract}


\section{Introduction}
\label{intro}
 Wolf-Rayet (WR) stars feature powerful high-velocity winds ($\dot{M}_{WR}\approx 10^{-5}~M_{\odot}$ yr$^{-1}$, $v\approx 10^3$ km s$^{-1}$), in which wide and intense helium and CNO-group emission lines are formed.
 %\deleted{ as well as lines the CNO group}. 
 According to modern concepts, massive WR stars 
 %\deleted{of the first type of Galaxy population} 
 are the naked (mostly helium) cores of initially massive Population I stars that lost their hydrogen envelopes either due to stellar-wind mass loss 
 %\deleted{in the form of a stellar wind} 
 \citep{Conti1975}, or due to the mass exchange in close binary systems \citep{Paczynski1973}. 
 %\deleted{Thus, the 'classical' WR stars are at a late stage of evolution}. 
 Collapses of carbon-oxygen nuclei of WR stars lead to type Ib/c supernova explosions and the formation of relativistic compact objects (e.g. \citealp{Dessart2020}).

To understand the physics and evolution of WR stars it is important to know their masses, radii, and mass-loss rates. The role of the WR mass loss is especially important in evolutionary calculations of the formation of merging binary relativistic objects 
%\deleted{that merge in a time shorter than the age of the Universe with the formation of observed bursts} 
- progenitors of the observed LIGO-Virgo gravitational wave sources \citep{Tutukov2020}.

The masses and radii of the 'cores' of WR stars are determined from the analysis of the radial velocity curves and the light curves of close binary WR+OB systems \citep{Cherepashchuk1984, Antokhin2007}. To determine the WR mass-loss rates, %\deleted{of $\dot{M}_{WR}$ WR stars}, 
researchers use spectroscopic methods for analyzing emission line profiles, methods based on the analysis of thermal radio emission or IR radiation from WR (single or in binary), or methods that analyze the orbital variable linear polarization of binary radiation from WR+OB systems. %\pk{This phrase is unclear}. 
However, spectroscopic methods for estimating $\dot{M}_{WR}$ are model-dependent. The values of $\dot{M}_{WR}$ estimated from the analysis of thermal radio emission or IR radiation from WR stars are burdened by the effects of stellar wind clumping. The most reliable method for estimating $\dot{M}_{WR}$ is the dynamic method based on the analysis of the secular change in the orbital period of a WR+OB system. For the first time, the dynamic method was applied by \cite{Khaliullin1974} who discovered the secular increase in the orbital period of the eclipsing binary system V444 Cyg (WN5+O6, $P_{orb} \approx 4^d.2$). Assuming that the orbital period change $\dot{P}$ is caused by the high-velocity radial wind  from  the WR star (the Jeans mass-loss mode), he gave an estimate $\dot{M}_{WR}=10^{-5}~M_{\odot}$ yr$^{-1}$. Later observations enabled %\deleted{allowed us to} 
improving the value of $\dot{P}$ to obtain $\dot{M}_{WR}= 7\cdot10^{-6}~M_{\odot}$ yr$^{-1}$, roughly consistent with the value of $\dot{M}_{WR}= 2.4\cdot10^{-5}~M_{\odot}$ yr$^{-1}$ found for V444 Cyg from the analysis of its radio and IR fluxes \citep{Prinja1990,Howarth1992}. %\deleted{Due to the fact that} 
Due to the effect of the WR wind clumping the estimates of $\dot{M}_{WR}$ by radio and IR fluxes are overestimated by factor 3-4. %\deleted{At the same time} 
In contrast, the estimate $\dot{M}_{WR} = (6-7)\cdot10^{-6}~M_{\odot}$ yr$^{-1}$ as inferred from  the orbital period change and the orbital variability of the system's emission polarization 
\citep{1988ApJ...330..286S}
does not depend on the WR wind structure and therefore is the most reliable.

%\pk{end of PK's editing}

%\pk{Below I didn't marked my changes anymore - too many...}

The search for changes in the orbital period from the analysis of the light curves of WR+OB systems is very promising for dynamic estimates of  $\dot{M}_{WR}$. Unfortunately, the number of eclipsing WR+OB binaries is very low: in the Northern sky, only V444 Cyg, CX Cep, CQ Cep and CV Ser can be observed. At the same time, there are several dozen non-eclipsing spectral-binary WR+OB systems  \citep{van2001viith}, in which the orbital period change $\dot{P}$ can be probed spectroscopically using the radial velocity curves of the components at different epochs. %Since the initial epochs of spectral observations of many WR+OB systems date back to 1940-1950, it is hoped that over the past 70-80 years, spectral observations will make it possible to notice changes in their orbital periods or, at least, with the help of new spectral observations, to implement another epoch of observations so that in the future the detection of a secular change in the orbital period of such WR+OB systems became possible.

In this paper, we present the results of spectral observations of two short-period WR+OB binary systems, CQ Cep and CX Cep. 
In Section 2, we overview the existing spectroscopic observations on these binaries. In Section 3, we describe our observations of CQ Cep and CX Cep. In Section 4, we discuss spectroscopic features of CQ Cep, the search for spectral traces of the OB-companion, and give a new estimate of the secular decrease of the orbital period of CQ Cep found by the spectroscopic method. In Section 5, we describe the spectroscopy of CX Cep, determine the binary system parameters and give a new estimate of secular increase in the orbital period of CX Cep. In Section 6, we report on searches for orbital period changes in CQ Cep and CX Cep. Section 7 presents the discussion of the results and estimates of mass-loss rates, and Section 8 is the conclusion. In Appendix, we describe the effect of the finite sizes of stars with stellar-wind mass loss in close binary systems on their orbital evolution.

%The application of the method of searching for changes in the orbital period  by the radial velocity curves made it possible to detect  system, which is consistent with $\dot{P}$, previously suspected from photometric data. 
%Due to the need to refer to archival data, in the section "Overview of CQ Cep and CX Cep spectroscopic studies"\ we provide a detailed description and analysis of them. 
%A secular elongation of the orbital period was detected in the CX Cep system by the spectroscopic method. The masses of the components and the sizes of their orbits are determined, and estimates of the mass loss rates for WR and OB satellite stars are given. 

\section{Overview of CQ Cep and CX Cep spectroscopic studies}
\label{overview}

\subsection{CQ Cep}

The spectral binarity of the star HD 214419 = CQ Cep was first reported by \cite{Mclaughlin1941}. \cite{Hiltner1944} presented and analyzed 70 spectrograms obtained at Ann Arbor on the 82-inch reflector of the McDonald Observatory %were analyzed in detail. 
He also gave the first reliable estimate of the orbital period: $P = 1.6410^d$. The radial velocity curves obtained from the emission lines $\lambda$4058 \ion{N}{iv} and $\lambda$4686 \ion{He}{ii} showed different behavior: while the radial velocity curve for $\lambda$4058 \ion{N}{iv} has an almost sinusoidal shape, that of $\lambda$4686 \ion{He}{ii} suggests a significant orbital eccentricity (0.35 according to Hiltner's graphical solution).

Among the early CQ Cep spectroscopy, observations made in 1951-1952 at the Mount Wilson Observatory with the 60-inch telescope \citep{Bappu1977} can also be noted. In this paper, the radial velocities from the lines $\lambda$4058 \ion{N}{iv} and $\lambda$4686 \ion{He}{ii} were given. 
%depending on the phase of the orbital period. 
The resulting elements turned out to be close to those obtained by \cite{Hiltner1944}.

\cite{Niemela1980} published the radial velocity curves of several binary stars with WR components. For CQ Cep, the radial velocity curve from the emission line $\lambda$4058 \ion{N}{iv} and the absorption $\lambda$4686 \ion{He}{ii} 
%pposite to it 
were presented and estimates of the semi-amplitude of the radial velocities, gamma-velocities and masses of the components were given. Using the radial velocity amplitudes, Niemela estimated the masses of the binary components: $M_{OB}\cdot\sin^3i= 19~M_\odot$, $M_{WR}\cdot\sin^3i = 23~M_\odot$ ($q=M_{WR}/M_{OB} = 1.21$).

\cite{Leung1983} presented the radial velocity curves constructed from a number of emission and absorption lines based on 26 spectrograms obtained in September 1978 with the 90-cm telescope of the Kitt Peak National Observatory. The positions of the Balmer hydrogen absorption lines, in addition to the systematic blue shift, changed in phase with the WR emission, indicating their origin in the WR atmosphere and not in the invisible companion star.

\cite{Stickland1984} analyzed observations of CQ Cep in a wide spectral range. These authors studied the UV spectra obtained by the IUE satellite, UBVJKL photometry, and using the original plates analyzed by \citet{Hiltner1944}, revisited the radial velocities constructed from different lines.
%were re-measured along a number of lines. 
They determined the spectral class of the WR star as WN7. No reliable traces of the secondary component were found. The interstellar lines studied by the high-resolution UV spectra made it possible to reliably attribute CQ Cep to the young star cluster Cep OB1.

The CQ Cep system was observed spectroscopically with the 6-m telescope of the Special Astrophysical Observatory in 1981 and 1982 \citep{Kartasheva1985}. A total of 16 spectra were obtained. Based on these data, radial velocities were constructed using the $\lambda$4058 \ion{N}{iv} emission and other lines (both emission and absorption); the radial velocity curves were constructed using the orbital elements from  \cite{Khaliullin1970}, and the masses of the components and the orbital sizes were calculated assuming zero orbital eccentricity.

The results of spectrophotometric and spectroscopic observations of CQ Cep in the range $\lambda\lambda$3800-6800\AA\ were presented in \cite{Shylaja1986}. In addition, spectrograms obtained by Bappu in 1951-1952 \citep{Bappu1977} were analyzed. Except for $\lambda$4603 \ion{N}{v}, the emissions fluxes showed an increase in the brightness minima. Different radial velocity curves 
%of radiation and absorption 
yielded different orbital solutions with a common positive shift (the gamma-velocity). The influence of geometric effects on the flux variations in the $\lambda$4058 \ion{N}{iv} line was noted. No traces of the O-companion were found.

Spectroscopic CCD observations were first presented by \cite{Marchenko1995}. These observations were carried out in 1986-1994 at the Mont M{\'e}gantic (Canada) and San Pedro Martir (Mexico) observatories. All the most prominent emission details in the spectrum varied in phase with WR emissions. A careful analysis enabled the authors to identify several weak absorption lines associated with the O-star. Firstly, the authors noted a complex variability of the blend profile at $\lambda\approx$4100\AA, which they decomposed into a relatively strong and wide \ion{He}{ii}$\lambda$4100 emission superposed on three weak \ion{Si}{iv} $\lambda$4089, $\lambda$4116 and H${\delta}$ absorption lines. In addition, \ion{He}{i} ($\lambda$4471, $\lambda$5876) and \ion{He}{ii} ($\lambda$5412) absorption lines were found in counter-phase to emission variations. Using these lines, the radial velocity curves were constructed. The \ion{Si}{iv} $\lambda$4116 radial velocity variations obtained by a sinusoidal fit folded with the orbital period was found to have the smallest semi-amplitude in comparison with all others because the \ion{Si}{iv} lines were not taken into account when calculating the orbital elements. Only the  \ion{N}{iv} $\lambda$4058 line was used to search for the WR orbital elements.
Assuming the most likely orbital inclination angle $i=65^o-78^o$, the authors give the following possible ranges of masses and sizes of the binary components: $M_{WR} = 15-19~M_{\odot}$, $M_O = 18-23~M_{\odot}$, $R_{WR} = 2-10~R_{\odot}$ $R_O < 10~R_{\odot}$. The mass ratio is $q = 0.87\pm0.04$. Also, based on the Beals method \citep{Beals1944}, the luminosity ratio  was estimated: $L_{WR}/L_{O} = 1.3\pm0.2$. The WR spectral class  is between WN6 and WN7, somewhat closer to the first. The spectrum of the companion star can be classified as O8-O8.5 of luminosity class V-III or as O8.5-O9.5 of luminosity class II-I.
By comparing the depths of the \ion{He}{ii} Pickering absorption lines related to the WR star, an increase in $\approx 15-20$\% of the lines superimposed on the hydrogen Balmer series was discovered. From the analysis of phase variations of equivalent widths and profiles of emission lines in the CQ Cep spectrum, as well as based on measurements of its X-ray luminosity by the ROSAT satellite \citep{Pollock1995}, the authors concluded that no strong stellar wind collision region  but rather a complex interaction of gas flows exist in the system.

The problem of visibility of the O-star lines in the CQ Cep spectrum and the  determination of the luminosity ratio $L_{WR}/L_{O}$ is discussed in \citet{Kartasheva1996}. Using the \citet{Beals1944} method, she calculated the components' luminosity ratio: $L_{WR}/L_O = 1.81$. After taking into account the probable triplicity of the system, in a later paper \citep{Kartasheva2000} this ratio was redetermined to be $L_{WR}/L_O = 2.7$.

High-precision spectropolarimetric observations of CQ Cep were obtained with the 4.2-m William Herschel Telescope (WHT) by \cite{Harries1997}. The radial velocity semi-amplitude $K_{WR} = 290\pm1$ km s$^{-1}$ and systematic velocity $\gamma_{WR} = -72\pm1$ km s$^{-1}$ were obtained using the \ion{N}{iv}$\lambda$4058 line, which is the best spectral diagnostic of the WR star orbital motion in this system, in good agreement with the results of the previous spectroscopic studies. Those authors identified the \ion{He}{i} $\lambda$4143 line as a spectral feature associated with the O-star. The semi-amplitude of its orbital variation was found to be $\gamma_O$ = 360$\pm$18 km s$^{-1}$ implying the mass ratio $q = M_{WR}/M_{O} = 1.24\pm0.16$ coinciding with the result of \cite{Niemela1980}.
Spectropolarimetric measurements combined with published photopolarimetric data allowed improving the inclination of the orbit ($i= 82^{\circ} .0\pm0^{\circ}.5$). It was noted that the orbital inclinations derived from the photometric analysis of eclipsing WR+O binaries are systematically lower than those calculated from polarimetric measurements.

%%%%%%%%%%% end of 2d PK's editing %%%%%%%%%%%

\subsection{CX Cep}

%As CX Cep is fainter than CQ Cep, significantly less work was performed with its spectroscopic measurements.

CX Cep  was identified as a WR star at the Linder McCormick Observatory in 1945 among six new objects, which, when observed with an objective prism, showed strong emission lines \citep{Vyssotsky1945}. The conclusion about the spectral binarity of the star with the observed details of both components (SB2), was made by \cite{Hiltner1946}. He also determined the spectral class of the star WR as WN5.

The earliest extensive photometric and spectral observations of CX Cep were reported by \cite{Hiltner1948}. He presented the radial velocity curve obtained from the bright $\lambda$4686 \ion{He}{ii} emission line using 70 spectrograms and the photoelectric light curve. The spectra were obtained from October 1945 to September 1947, photometric measurements were carried out in August and September 1947. Observations were carried out at the McDonald Observatory. The orbital period ($P = 2.1267^d$) and semi-amplitude of the WR radial velocity ($K_{WR} = 290$ km s$^{-1}$) were estimated for the first time. The light curve showed the form characteristic for an eclipsing binary system with an almost zero eccentricity: a symmetric deep ($0.13^m$) primary minimum and a weaker ($0.04^m$) symmetric secondary minimum. 
%The depth of the primary and secondary minimum is about 0.13$^m$ and 0.04$^m$. 
The minima are wide due to the extended atmosphere of one of the components.

%In the General Catalog of Variable Stars (\cite{Samus2017}) this binary star is designated CX Cep.

Using the radial velocity curve and orbital period obtained by Hiltner, \cite{Bracher1966} estimated the orbital parameters of the CX Cep binary system. The semi-amplitude of the radial velocity curve of the WR star was found to be 302 km s$^{-1}$, the mass function $f(M_{WR}) \approx 6~M_\odot$, $a\cdot\sin{i}\approx 8.8 \cdot 10^6~\mbox{km} \approx 12.6~R_\odot$, $v_\gamma\approx +12$ km s$^{-1}$.

After \cite{Hiltner1948}, no detailed photometric or spectral observations of CX Cep were published until the early 1980s. In 1981, a series of papers by Massey and Conti devoted to spectroscopic studies of WR stars with absorption lines was published. In September 1978 and September-October 1979, the authors obtained 8 CX Cep spectra in the $\lambda\lambda$3700-4900\AA\ range \citep{Massey1981}. The absorption in the $\lambda$4200 \ion{He}{ii} line observed on top of the emission and the weak absorption detail $\lambda$4471 \ion{He}{i} related to the secondary star suggested its spectral class as O8V. The WR star class may be earlier: in the catalogues \cite{van1981sixth} and \cite{van2001viith}, the spectral class WN4 is referred to this work. Using Wilson's method \citep{Wilson1941}, %the binary components 
the mass ratio was found to be $M_O/M_{WR}=~2.34\pm0.21$. The radial velocity semi-amplitudes were estimated to be $K_{WR}~\approx 300~\mbox{km s}^{-1}$, $K_O~\approx 130~\mbox{km s}^{-1}$. Taking into account the significant orbital inclination ($i\geq50^\circ$), the masses of the binary components were found to fall in the range $M_{WR}\approx 5-12~M_\odot$, $M_{O}\approx 12-27~M_\odot$.

\cite{Lewis1993} analyzed 60 CX Cep spectra with a resolution of 5\AA. They measured the semi-amplitudes of radial velocities $K_O = 240\pm 8$ km s$^{-1}$, $K_{WR} = 340\pm 10$ km s$^{-1}$, corresponding to the masses $M_O\cdot\sin^3i = 25.2\pm 1.9~M_\odot$, $M_{WR}\cdot\sin^3i = 17.8 \pm 1.4~M_\odot$. The WR spectral class was redefined to be O5. The authors checked for additional periods in the $\lambda$4686 \ion{He}{ii} radial velocity variability and found a period corresponding to one third of the main period, possibly due to tidal distortions in an extended emission envelope. Phase-dependent variations of the equivalent width of the absorption line $\lambda$3888 \ion{He}{i} were also noted. The authors proposed a qualitative model of the CX Cep system which includes a WR star with an extended stratified expanding shell, an O star with a much weaker wind, and a shocked wind collision region producing an excess in the \ion{He}{i} and \ion{He}{ii} emissions.

\section{Observations}
\label{obs}

Our spectroscopic observations of CQ Cep and CX Cep were carried out in the period from November 2020 to January 2023 on the 2.5-meter telescope of the Caucasian Mountain Observatory of Moscow State University \citep{Shatsky2021} using a Transient Double-beam low-resolution Spectrograph (TDS). TDS registers the spectrum in the 'blue' ($\lambda\sim360-577$ nm, R = 1300) and 'red' ($\lambda\sim567-746$ nm, R = 2500) channels (see  \citealp{Potanin2020}) for a detailed description and characteristics of the spectrograph and data reduction).

A total of 199 CQ Cep spectra (70 nights) and 183 CX Cep spectra (71 nights) were taken and processed. %We publish the observation log separately \pk{??}. 
The processed data make it possible to obtain fairly phase-dense time series of radial velocities in the emission and absorption lines of various ions. The method of measuring radial velocities is described in detail below.

Strong and wide emission lines of ionized helium in the CX Cep spectrum are blended by counter-phase \ion{He}{ii} and \ion{H}{i} absorptions related to the O-star. To measure the positions of these emissions and plot the radial velocity curves, a technique based on combining the wings of the profiles was used. To do this, the profiles were cut at half intensity and combined with the averaged profile. %Performing 3-4 iterations of this algorithm results in a radial velocity curve with an average error of $\bar{\sigma}\sim$20-50 km s$^{-1}$ (depending on the range and intensity of the line). A further increase in the number of iterations does not lead to a significant improvement in accuracy. In Fig. \ref{HeII_align} the operation of this algorithm is presented on the example of the $\lambda$4686 \ion{He}{ii} line in the CX Cep spectrum.
In the CQ Cep spectra, \ion{He}{ii} lines demonstrate a more complex profile, which is a mixture of a wide P Cyg profile with a probable counter-phase emission. Due to this complexity, which is difficult to eliminate with our spectral resolution, we have not used the radial velocities measured by \ion{He}{ii} emissions to search for the orbital period in the case of CQ Cep. The \ion{He}{i} lines have also not been used to search for the period and calculate the orbital parameters.

%\begin{figure*}
   %\includegraphics[scale=0.38]{HeII_alingment_new.png}
   %\caption{The result of the algorithm for combining lines on the example of the $\lambda$4686 \ion{He}{ii} line in the CX Cep spectrum. The upper panels are the observed profiles of the spectral line and the result of their combination (on the left - in the form of a dependence of the normalized flux on the wavelength, on the right - in the form of a heat map giving a phase sweep). The lower left panel is the time series of the measured radial velocities, the lower right is the convolution of the time series with the calculated period, the color encodes the applied method (black is the LS spectrum, green is the stretched string method, blue is the LK method)}
   %\label{HeII_align}
%\end{figure*}

Weak absorption and emission lines and resolved blends of weak lines were measured by fitting with Gaussians (one for a single line or two for a blend of two lines). While the true profiles of the lines may strongly differ from the Gaussian form, the low spectral resolution enabled us to find the central line wavelengths %obtained for each spectrum adequately describe the line positions.
and to use them to construct radial velocity curves.

When calculating the lines positions in the 'blue' channel and its errors we took into account the systematic deviation of the calibration and the random deviation due to the displacement of the star relative to the slit center.

%Thus, using the sequence of profiles of a certain emission line, it is possible to obtain a time series of radial velocities and individual errors in their determination. 
A significant number of spectra and good coverage of the entire area of possible line positions make it possible to calculate the orbital period directly from spectral measurements. 
To search for periodicity, we have used: 1) the RMS spectrum of correspondence to an oscillation with a certain trial period - the so-called LS-spectrum \citep{Barning1963}, and 2) the Lafler-Kinman (LK) method \citep{Lafler1965}. The  use of two different methods  allows us to assess the degree of deviation of the radial velocity curve from the expected sinusoidal shape.
Table \ref{pers_CQCX} lists the elements calculated from the constructed radial velocity curves. In our analysis, we accepted the periods calculated by the LK method.

\begin{table}
    %\scriptsize
    \centering
    \caption{Periods calculated by CQ Cep and CX Cep lines}
    \begin{tabularx}{\columnwidth}{lcXXX}
\hline
{Line (CQ Cep)} & Type & MJD0 & P$_{LS},$ d & P$_{LK},$ d \\
\hline
$\lambda$4058 \ion{N}{iv}  & Emis.   & 59561.651643   & $1.641315$ & $1.641369$  \\
$\lambda$5204 \ion{N}{iv}  & Abs.    & 59561.651643   & $1.641256$ & $1.641236$  \\
$\lambda$5204 \ion{N}{iv}  & Emis.   & 59561.651643   & $1.641320$ & $1.641216$  \\
$\lambda$4603 \ion{N}{v}   & Abs.    & 59561.638205   & $1.641093$ & $1.641233$  \\
$\lambda$4603 \ion{N}{v}   & Emis.   & 59561.651643   & $1.641266$ & $1.641241$  \\
$\lambda$4619 \ion{N}{v}   & Abs.    & 59561.640118   & $1.641378$ & $1.641241$  \\
\\
\multirow{2}{*}{Median} &  & 59561.651643     & $1.641291$ & $1.641239$  \\
                       &   & $\pm00590$ & $\pm000089$ & $\pm000051$  \\
\\ 
\hline
{Line (CX Cep)} & Type & MJD0 & P$_{LS},$ d & P$_{LK},$ d \\
\hline                                                                             
H$_9$-H$_{11}$       & Abs.    & 59562.152118  & $2.126989$ & $2.128199$ \\  
$\lambda$4685 \ion{He}{ii}   & Emis.   & 59561.192653  & $2.127004$ & $2.127166$ \\  
$\lambda$4859 \ion{He}{ii}   & Emis.   & 59561.183098  & $2.126929$ & $2.126895$ \\
$\lambda$5411 \ion{He}{ii}   & Emis.   & 59561.019592  & $2.127003$ & $2.127153$ \\  
$\lambda$6560 \ion{He}{ii}   & Emis.   & 59561.235782  & $2.126999$ & $2.126756$ \\  
$\lambda$4057 \ion{N}{iv}    & Emis.   & 59561.135380  & $2.127086$ & $2.127144$ \\  
$\lambda$7115 \ion{N}{iv}    & Emis.   & 59561.139464  & $2.127120$ & $2.127075$ \\  
$\lambda$4603 \ion{N}{v}     & Emis.   & 59561.140042  & $2.127353$ & $2.126870$ \\
$\lambda$4619 \ion{N}{v}     & Emis.   & 59561.139071  & $2.127073$ & $2.127056$ \\
$\lambda$4944 \ion{N}{v}     & Emis.   & 59561.169246  & $2.126918$ & $2.127035$ \\
\\                            
\multirow{2}{*}{Median}   & & 59561.14        & $2.127004$  & $2.127066$  \\
      &                      & $\pm06$         & $\pm000118$ & $\pm000377$ \\
    \end{tabularx}
    \label{pers_CQCX}
\end{table}

\begin{figure*}
   \centering
   \includegraphics[scale=0.52]{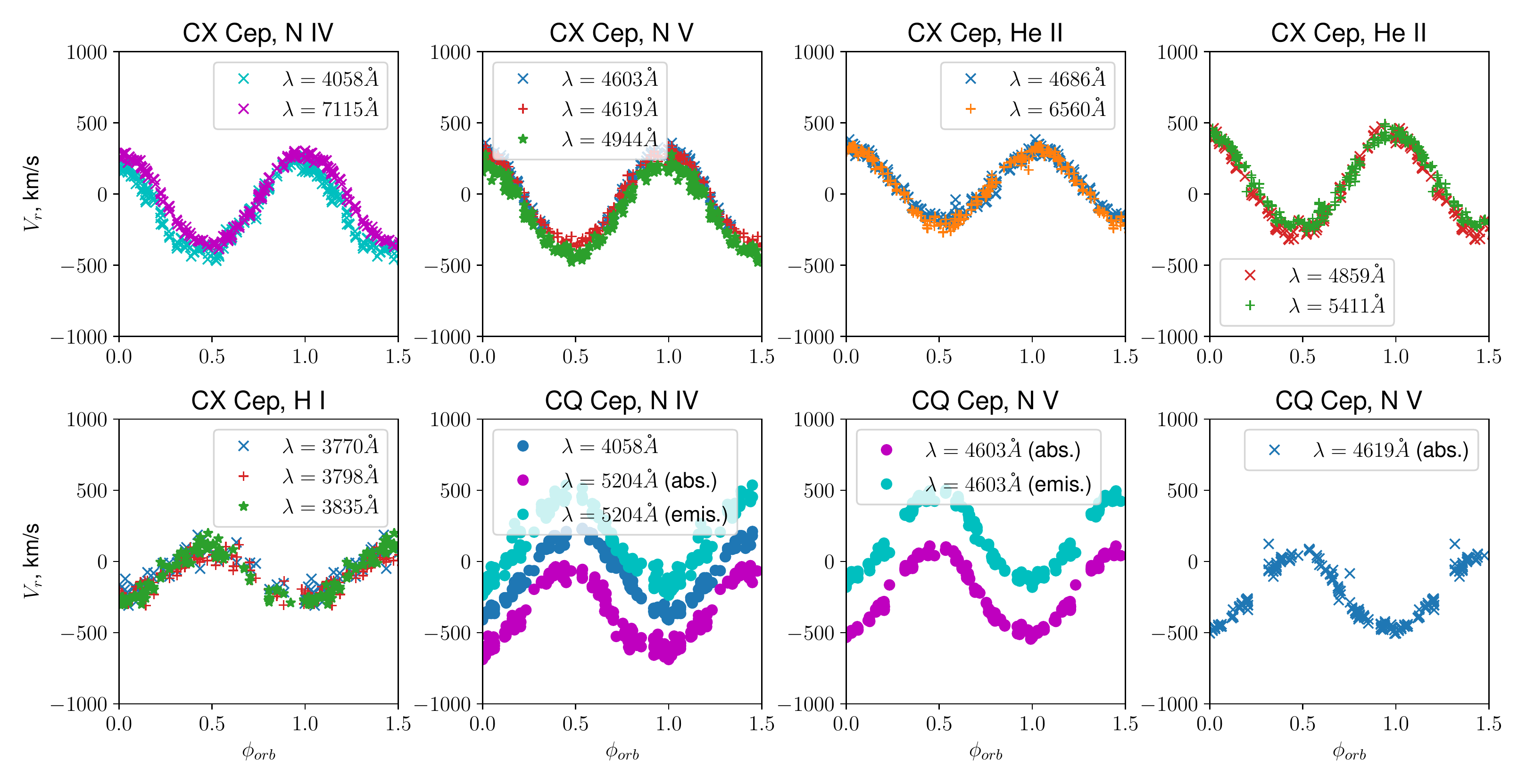}
   \caption{Radial velocity curves of CQ Cep and CX Cep constructed from different spectral lines.}
   \label{vel}
\end{figure*}

Fig. \ref{vel} shows the radial velocity curves folded with the found period for each object.

%We show the CQ Cep and CX Cep spectrograms separately in the form of spectral atlases, which represent a smoothed sweep along the orbital phase with a period calculated for each object. The normalized intensity is color coded, the scale for each panel is selected individually and shown on the right. All identified spectral features are marked in the atlases. The telluric and interstellar features are marked separately in green (the latter are signed).

\section{Spectroscopic features of the CQ Cep system. Search for OB-star traces in the spectrum.}

\begin{figure*}
   \centering
   \includegraphics[scale=0.54]{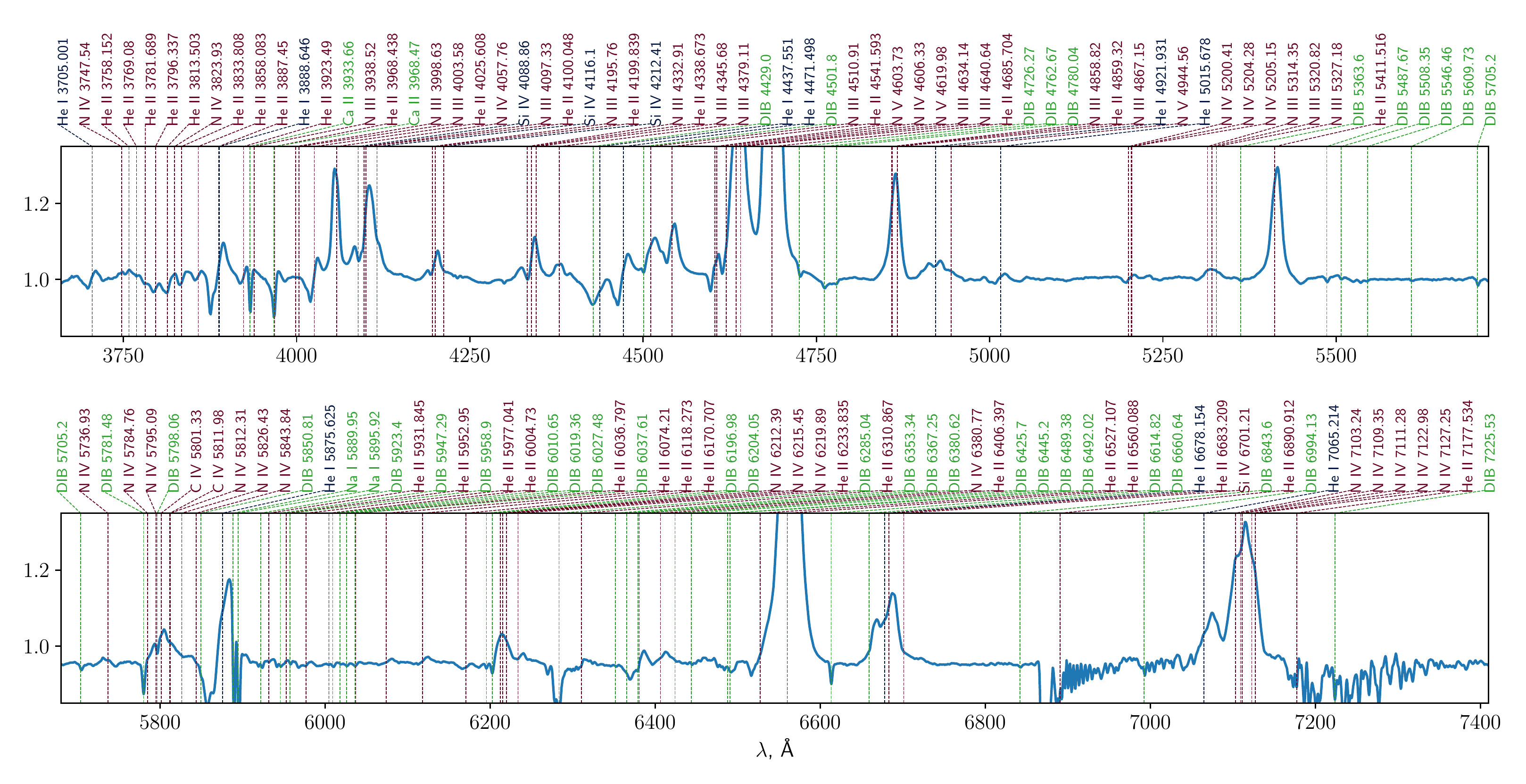}
   \caption{The averaged spectrum of CQ Cep. The most noticeable spectral details are marked. The interstellar lines are in green, absorption lines are in black and emission lines are in red.}
   \label{mean_CQ}
\end{figure*}

Fig. \ref{mean_CQ} displays the averaged (excluding radial velocity shifts) spectrum of CQ Cep.
The strongest details in the CQ Cep spectrum are \ion{He}{ii} emission lines: $\lambda$4686, $\lambda$4859, $\lambda$5411, $\lambda$6560. The emissions of \ion{N}{iv} $\lambda$4058, $\lambda$4606 and a blend near $\lambda$7115 \AA\ are also visible. There is a noticeable phase-stable \ion{N}{v} $\lambda$4603 detail with P Cyg profile. The \ion{N}{iii} and \ion{N}{iv} line intensities are comparable, which, considering the weakness of the \ion{N}{v} line, suggests the spectral class WN6 of the WR star.

The spectral features demonstrate a complex phase dependence. The \ion{N}{iv} lines noticeably change their relative intensities reaching maximum near the strongest red shift. The most prominent is the variability of neutral and ionized helium line profiles. The \ion{He}{ii} lines show type III P Cyg profiles  according to \cite{Beals1944}, with the intensity of the emission part and the depth of the absorption part being in the counter phase.

In the case of CQ Cep, as a rule, the radial velocity curve of the WR star is calculated using the \ion{N}{iv} $\lambda$4058 line. Additionally, the most contrast spectral features related to the \ion{N}{iv}, \ion{C}{iv} and \ion{N}{v} ions were measured. %Taking $K_{WR} = 297.0 \pm 1.2$ km s$^{-1}$ and $e = 0$, we can  estimate the size of the orbit and the mass function: $a_{WR}\sin{i} = 9.63 \pm 0.04 R_{\odot},~f_{WR}(M) =  \frac{M_O^3 \sin^3 i}{(M_{WR}+M_O)^2} = 4.46 \pm 0.05 M_{\odot}.$
Taking $K_{WR} = 297 \pm 9$ km s$^{-1}$ (from \ion{N}{iv} and \ion{N}{v} curves except \ion{N}{v} $\lambda4619$) and $e = 0$, we can estimate the size of the orbit and the mass function: $a_{WR}\sin{i} = 9.6 \pm 0.3 R_{\odot},~f_{WR}(M) =  \frac{M_O^3 \sin^3 i}{(M_{WR}+M_O)^2} = 4.46 \pm 0.27 M_{\odot}.$

The binary mass ratio is a critically important parameter to understand the orbital evolution of a close binary experiencing mass transfer.
%depends on the ratio of their masses, and, consequently, the further fate of the system. 
Unlike CQ Cep, in the CX Cep spectrum the absorption lines of the O-component are clearly visible and make it possible to obtain a reliable radial velocity curve. The poor visibility of the spectral traces of the secondary component affected the previous spectroscopic studies of this system. Some researchers detected weak absorption details in counter-phase offset relative to the bright emissions. 
%It is necessary to determine the $q = M_{WR}/M_{O}$ ratio.

%The rich spectral material % collected in this work 
%and the TDS resolution of the spectra obtained on TDS gave optimism that it would be an easy task to detect spectral details directly describing the motion of an O-star in the CQ Cep spectrum. However, in the end, this problem turned out to be the most time-consuming and controversial in this study.

\begin{figure*}
  \centering
  \includegraphics[scale=0.5]{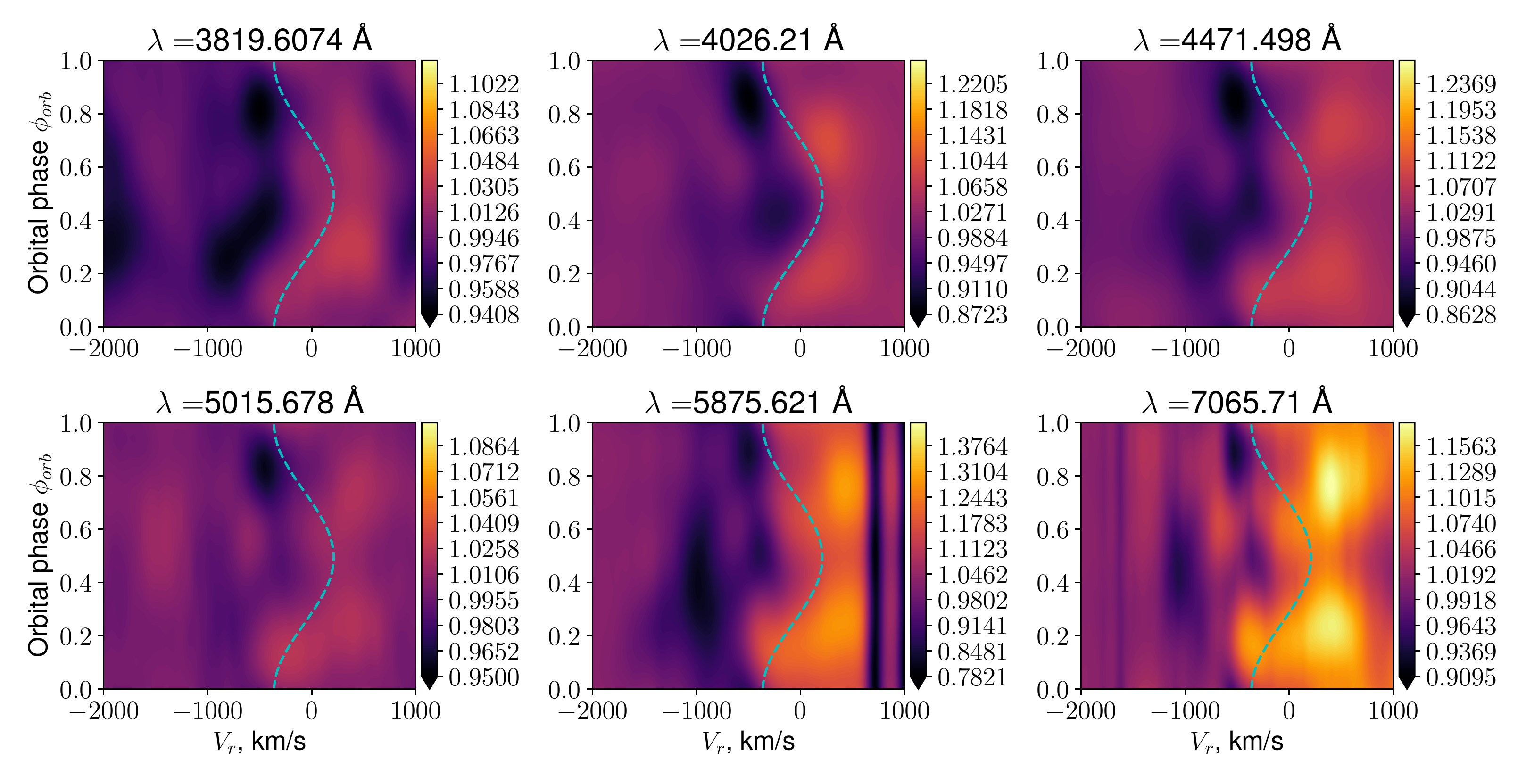}
  \caption{Profiles of six \ion{He}{i} lines, constructed in phase sweep from the CQ Cep TDS spectra. The dashed line shows a sine fit to the radial velocity curve in the \ion{N}{iv} $\lambda$4058 emission line.}
 \label{HeI_prof_CQ}
\end{figure*}

\ion{He}{i} absorption lines are most probably produced by the O-star. These lines were indeed detected by us, and they showed encouraging behavior: each line turned out to be a blend of two absorptions, one shifted synchronously in phase with the WR lines, and the second with a shift in the opposite direction. In Fig. \ref{HeI_prof_CQ} we show the phase-expanded profiles of several strongest \ion{He}{i} lines. The dashed line shows the sine fit to the radial velocity curve in the \ion{N}{iv} $\lambda$4058 line caused by the orbital motion of the WR star. The lines $\lambda$4024, $\lambda$5876 and $\lambda$7066 are blended with \ion{He}{ii} emissions. A strong blue shift of absorption is prominent. Comparison with the radial velocity curve shows that the right absorption component \ion{He}{i} (in-phase with the WR) does not reflect the motion of the WR star. We also state that the left absorption component \ion{He}{i} (out-of-phase with the WR) does not represent  the orbital motion of the OB star.
These absorptions may be associated with dense gas flows projected onto the photospheres of the stars. Another possibility is the influence of the WR radiation and the WR wind on the atmosphere of the OB star: if we assume that the OB star fills its Roche lobe and the mass ratio $q = M_{WR}/M_{OB}<1$, a counter-phase motion of parts of the OB star surface lying on the opposite sides from the barycenter is allowed. However, this does not explain the strong blue shift \ion{He}{i} lines.

\begin{figure*}
  \centering
  \includegraphics[scale=0.52]{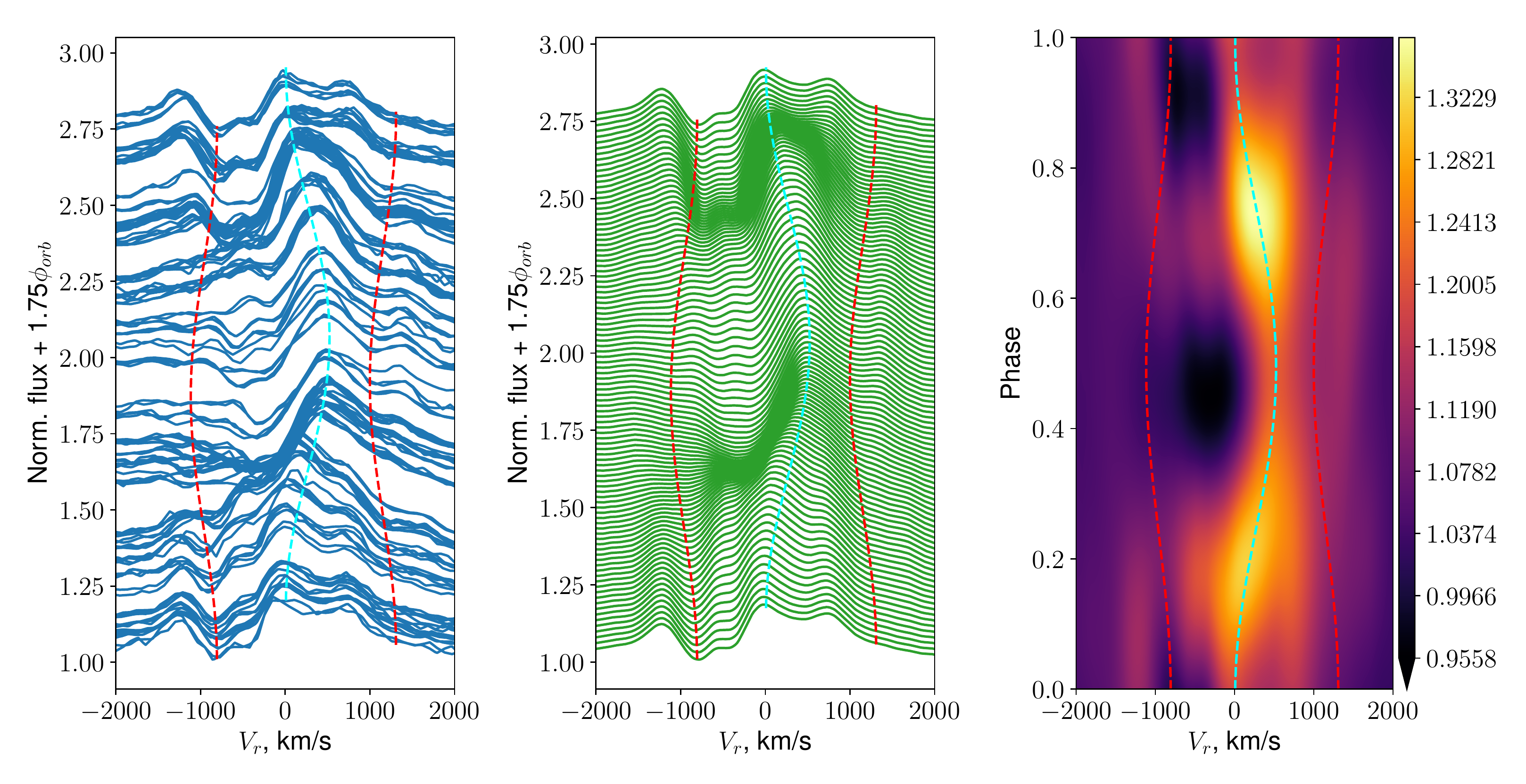}
  \caption{Blend profiles $\lambda$ = 4100\AA\ in the CQ Cep spectrum depending on the orbital phase. The left panel demonstrates the original spectra, the middle panel presents the smoothed spectra within  0.01 phase bins, the right panel is the same as the middle one in the form of a heat map. %\pk{What is it?} "A heat map (or heatmap) is a data visualization technique that shows magnitude of a phenomenon as color in two dimensions." 
  The estimated position of \ion{Si}{iv} $\lambda$4088, $\lambda$4116 absorptions (red) and the emission part of \ion{He}{ii} $\lambda$4100 (cyan) are shown.}
 \label{4100_prof_CQ}
\end{figure*}

Another possibility is to try to resolve the blend near $\lambda$= 4100\AA. In \cite{Marchenko1995}, the absorption lines \ion{Si}{iv} $\lambda$4088, $\lambda$4116 and H${\delta}$ were found. Fig. \ref{4100_prof_CQ} shows a part of the spectrum in the range $\lambda$4065-4135 \AA\ (in terms of velocities relative to \ion{He}{ii} $\lambda$ = 4100.048 \AA). The left panel shows the phase-ordered spectra with the signal-to-noise ratio $\ge$ 75 in the selected spectral range. The middle and right panel show the spectra averaged within 0.01 phase bins. The cyan dashed line marks the \ion{He}{ii} $\lambda$4100 emission from WR-star, the red lines 
%crosses and dashed lines 
show the probable \ion{Si}{iv} absorptions. 
The left absorption detail (the left red curve), identified as OB-star absorption line \ion{Si}{iv} $\lambda$4088, becomes distinguishable at phases with the highest red shifts. The right absorption detail (the right red curve), identified as OB-star absorption line \ion{Si}{iv} $\lambda$4116, becomes clearly visible 
%on top of the red wing of the bright emission blend shifting redwards (the middle of the left panel), i.e. 
at phases with the highest blue shifts. The semi-amplitude of the emission line radial velocity curve of the WR-star (the cyan line) is about 262 km s$^{-1}$. A joint analysis of both absorption lines related to one ion \ion{Si}{iv} (the red curves)  lead to a semi-amplitude of the OB-star about 155 km s$^{-1}$ suggesting the mass ratio 
$q \approx 0.6$. 

\begin{table}
    \centering
    \scriptsize
    \caption{Semi-amplitudes of radial velocities and $\gamma$-velocities for various spectral lines in the spectrum of CQ Cep reported by different authors.}
    \begin{tabularx}{\columnwidth}{XXcXXc}
        \hline
        mean MJD & Line & Type & $K$, km s$^{-1}$ & $\gamma$, km s$^{-1}$ & Ref.$^a$ \\
        \hline
        31044  & \ion{N}{iv}, $\lambda$4058     & Emis.       & 295 & -75                        & H44 \\
               & \ion{He}{ii}, $\lambda$4686    & Emis.       & 165 & 137                        &   \\
        34013  & \ion{N}{iv}, $\lambda$4058     & Emis.       & 295.0 & -61.6                    & B77 \\
               & \ion{He}{ii}, $\lambda$4686    & Emis.       & 148.2 & 117.5                    &   \\
        ?      & \ion{N}{iv}, $\lambda$4058     & Emis.       & 285 & -85                        & N80 \\
               & \ion{He}{i}, \ion{He}{ii}             & Abs.        & 340 & -220                       &  \\
        43770  & \ion{N}{iv}, $\lambda$4058     & Emis.       & $310\pm7 $ & $-60 \pm5 $        & L83 \\
               & \ion{N}{v}, $\lambda$4604      & Emis.       & $280\pm10$ & $ 153\pm6 $        &   \\
               & \ion{N}{v}, $\lambda$4604      & Abs.        & $299\pm8 $ & $-220\pm9 $        &   \\
               & \ion{N}{v}, $\lambda$4620      & Abs.        & $250\pm12$ & $-233\pm8 $        &   \\
               & \ion{He}{ii}, $\lambda$4542    & Emis.       & $312\pm13$ & $+148\pm8 $        &   \\
               & \ion{He}{ii}, $\lambda$4542    & Abs.        & $276\pm10$ & $-303\pm6 $        &   \\
               & \ion{H}{i}, $\lambda$3835      & Abs.        & $257\pm18$ & $-261\pm12$        &   \\
        31044  & \ion{N}{iv}, $\lambda$4058     & Emis.       & $297.4\pm2.7 $ & $ -53.4\pm2.3 $ & S84 \\
               & \ion{He}{ii}, $\lambda$4686    & Emis.       & $164.8\pm4.3 $ & $ 133.8\pm3.3 $ &  \\
               & \ion{He}{ii}, $\lambda$4859    & Emis.       & $235.8\pm23.7$ & $ 198.2\pm12.3$ &  \\
               & N III, $\lambda$4640    & Emis.       & $127.0\pm6.0 $ & $-114.5\pm4.9 $ &  \\
        44968  & \ion{N}{iv}, $\lambda$4058     & Emis.       & $290.9\pm11.8$ & $ -55.4\pm7.7 $  & KS85 \\
               & \ion{He}{i}, \ion{He}{ii}, \ion{H}{i}        & Abs.        & $179.0\pm6.4 $ & $   1.2\pm4,2 $  &  \\
               & \ion{He}{ii}, $\lambda$4859    & Emis.       & $221.9\pm12.2$ & $ 196.0\pm11.4$  & S86 \\
               & \ion{He}{ii}, $\lambda$6562    & Emis.       & $212.0\pm12.8$ & $ 206.0\pm10.2$ &  \\
               & N III, $\lambda$4640    & Emis.       & $127.8\pm10.4$ & $-114.0\pm8.2 $ &  \\
               & \ion{N}{v}, $\lambda$4604      & Emis.       & $286.0\pm6.4 $ & $ 159.6\pm7.2 $ &  \\
               & \ion{N}{v}, $\lambda$4604      & Abs.        & $307.5\pm7.3 $ & $-216.1\pm5.8 $ &  \\
        47167  & \ion{N}{iv}, $\lambda$5736.94  & Emis.       & $290\pm 5 $ & $-87 \pm3 $       & U90 \\
               &  \ion{C}{iv}, $\lambda$5811.98 & Emis.       & $328\pm 18$ & $-96\pm12$        &  \\
               &  \ion{C}{iv}, $\lambda$5801.33 & Emis.       & $362\pm 29$ & $ -50 \pm19$        &  \\
               &  N II, $\lambda$5321.72 & Emis.       & $148\pm 18$ & $ -75 \pm12$        &  \\
               & \ion{He}{ii}, $\lambda$5411.52 & Emis.       & $277\pm 13$ & $ 140\pm8 $       &  \\
               & \ion{He}{ii}, $\lambda$5411.52 & Abs.        & $260\pm 50$ & $-445\pm32$    &  \\
               &  \ion{He}{i}, $\lambda$5875.62 & Abs.        & $196\pm 49$ & $ -480\pm32$    &  \\
               &  \ion{N}{iv}, $\lambda$5203.21 & Abs.        & $256\pm 18$ & $ -280\pm12$    &  \\
        48447  & \ion{N}{iv}, $\lambda$4058     & Emis.       & $303.2\pm2.4 $ & $-58.5\pm1.7 $  & M95 \\
               & \ion{He}{i}, \ion{He}{ii}, \ion{H}{i}        & Abs.        & $263.5\pm10.6$ & $-57.6\pm18.0$  &  \\
               & \ion{Si}{iv}, $\lambda$4116    & Abs.        & $197.5\pm15.7$ & $-60.0\pm11.1$ &  \\
        49940  & \ion{N}{iv}, $\lambda$4058     & Emis.       & $290\pm1 $ & $-72\pm1 $          & HH97 \\
               & \ion{He}{i}, $\lambda$4143     & Abs.        & $360\pm18$ & $-34\pm14$          & \\
        59504  & \ion{N}{iv}, $\lambda$4057 	 & Emis.       & $286\pm 4$ & $ -55\pm7$   & $\dag$ \\
               & \ion{N}{iv}, $\lambda$5204     & Abs.        & $288\pm 5$ & $-357\pm6$  &    \\
               & \ion{N}{iv}, $\lambda$5204     & Emis.       & $311\pm 6$ & $ 152\pm6$  &    \\
               & \ion{N}{iv}, $\lambda$5737     & Emis.       & $289\pm 4$ & $ -84\pm7$   &    \\
               & \ion{N}{v}, $\lambda$4603      & Abs.        & $299\pm 3$ & $-228\pm4$  &    \\
               & \ion{N}{v}, $\lambda$4603      & Emis.       & $308\pm 4$ & $ 166\pm4$   &    \\
               & \ion{N}{v}, $\lambda$4619      & Abs.        & $272\pm 4$ & $-220\pm5$  &    \\
               & \ion{C}{iv} $\lambda$5812      & Emis.       & $303\pm 6$ & $ -78\pm6$   &    \\
    \hline
    \multicolumn{6}{c}{$^a$H44=\cite{Hiltner1944}, 
    B77=\cite{Bappu1977}, N80=\cite{Niemela1980},}\\ 
    \multicolumn{6}{c}{
    L83=\cite{Leung1983}, S84=\cite{Stickland1984}, KS85=\cite{Kartasheva1985},}\\ \multicolumn{6}{c}{
    S86=\cite{Shylaja1986}, U90=\cite{Underhill1990}, M95=\cite{Marchenko1995}, }\\
    \multicolumn{6}{c}{
    HH97=\cite{Harries1997},
    $\dag$ = present paper}
    \end{tabularx}
    \label{vel_compar_CQ}
\end{table}

The estimated mass ratio differs from previous estimates: 1.21 \citep{Niemela1980}, 1.24$\pm$0.16 \citep{Harries1997}. The mass ratio $q=0.584$ reported by \cite{Kartasheva1985} is close to our result, but later they changed their estimate to 0.83$\pm$0.07, which is close to $0.87\pm0.04$ found by \cite{Marchenko1995}. In Table \ref{vel_compar_CQ} we summarize available radial velocity semi-amplitude and $\gamma$-velocity measurements in various spectral lines in CQ Cep. Clearly, the large dispersion in $\gamma$-velocities for different lines makes it difficult to accurately determine the radial velocity amplitude by using several lines of different ions.

\begin{figure}
   \centering
   \includegraphics[scale=0.45]{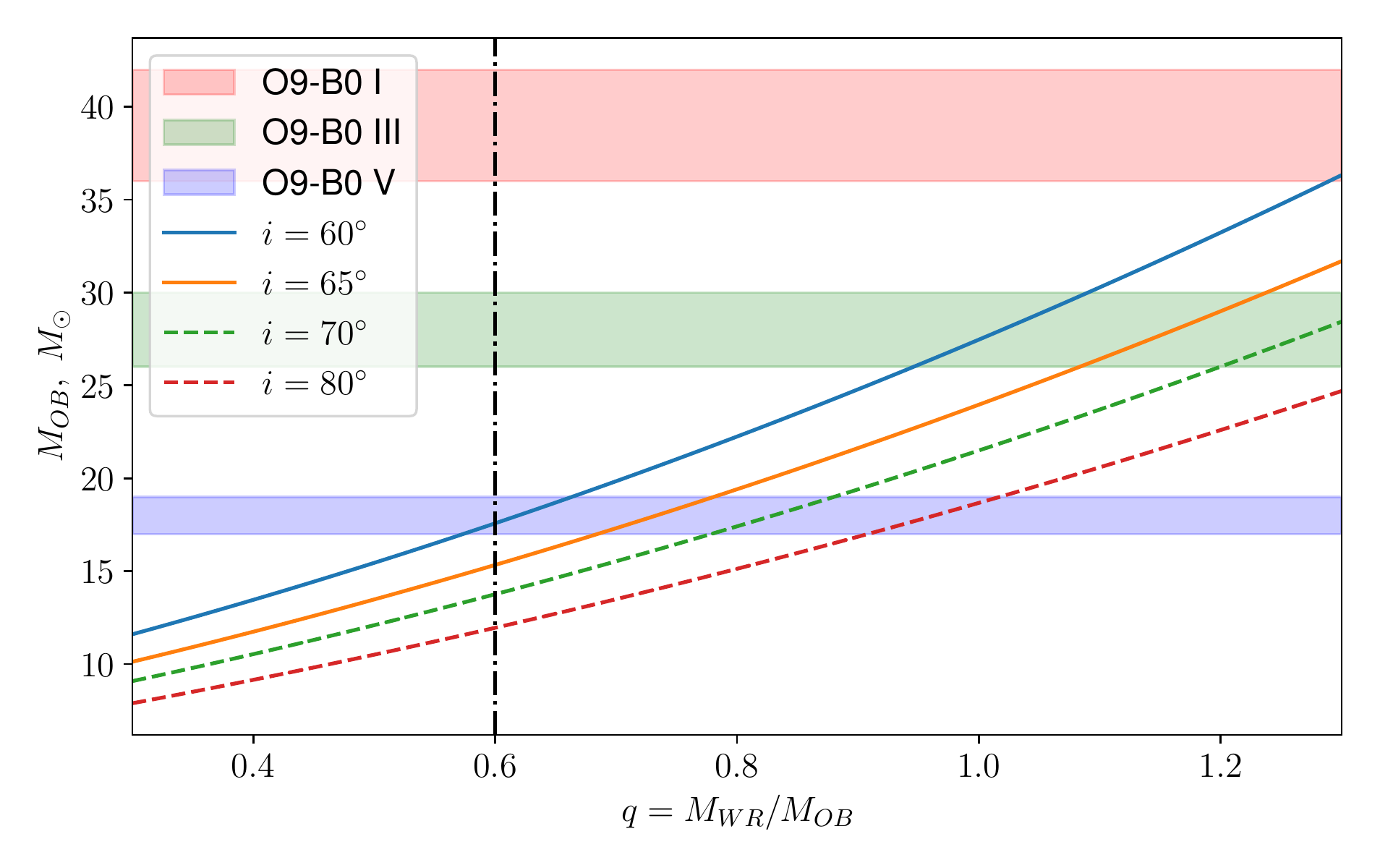}%
   %\caption{The OB-star mass in the CQ Cep system for different $q$ and $i$.}
   \caption{The OB-star mass in the CQ Cep system for different $q$ and $i$ with given $f_{WR}$: $M_{OB} (q, \sin{i}) = \frac{f_{WR}}{\sin^3{i}} (1+q)^2$. }
   \label{M_OB_CQ}
\end{figure}

Fig. \ref{M_OB_CQ} shows the dependence of the mass $M_{OB}$ on the mass ratio $q = M_{WR}/M_{OB}$ given the found value of the mass function $f_{WR}(M) = 4.46~M_{\odot}$ for different binary inclinations angles $i$. The colored horizontal strips correspond to mass intervals of O9-B0 I-III-V stars according to model calculations by \cite{Weidner2010}. 
The estimated mass ratio $q = M_{WR}/M_{OB}\approx 0.6$ suggests an O9.5-B0 V star with a mass of 18 $M_{\odot}$ (the bottom violet strip) and an orbital inclination of  $60^{\circ}$ as the secondary companion to the WR star in CQ Cep. Then $M_{WR}$ should be  about 11 $M_{\odot}$ 
%The mass ratio $q = 0.8$ and inclination $i = 70^{\circ}$ would give a WR mass of $\sim 14 M_{\odot}$  corresponding to WN6 spectral type  %in early papers  (for example, in citealp{hamann2006galactic}), or 
similar to a WN6 star  with a mass of 12 $M_\odot$ in the Potsdam Wolf-Rayet models \citep{todt2015potsdam}. %and therefore we consider the values $q = 0.8$ and $14~M_\odot$ exceeded.
Considering the range of the orbital inclinations $i=60-65^\circ$ as inferred from the photometric analysis of CQ Cep \citep{1982SvA....26..569L},  
the plausible mass of the WR star in CQ Cep is $M_{WR} \sim 11-14 M_{\odot}$. Therefore, the mass ratio $q\sim 0.6$ is consistent with our spectroscopic measurements and the binary inclination angle determined from the photometry of CQ Cep. %The values of $q$ in the range 0.8-1.0 would correspond to larger $i$, for example, such as obtained for CQ Cep by the polarimetric method, but this method is known to overestimate the binary inclinations.

\section{Spectroscopic features of the CX Cep system. Determination of the binary system parameters}

\begin{figure*}
   \centering
   \includegraphics[scale=0.54]{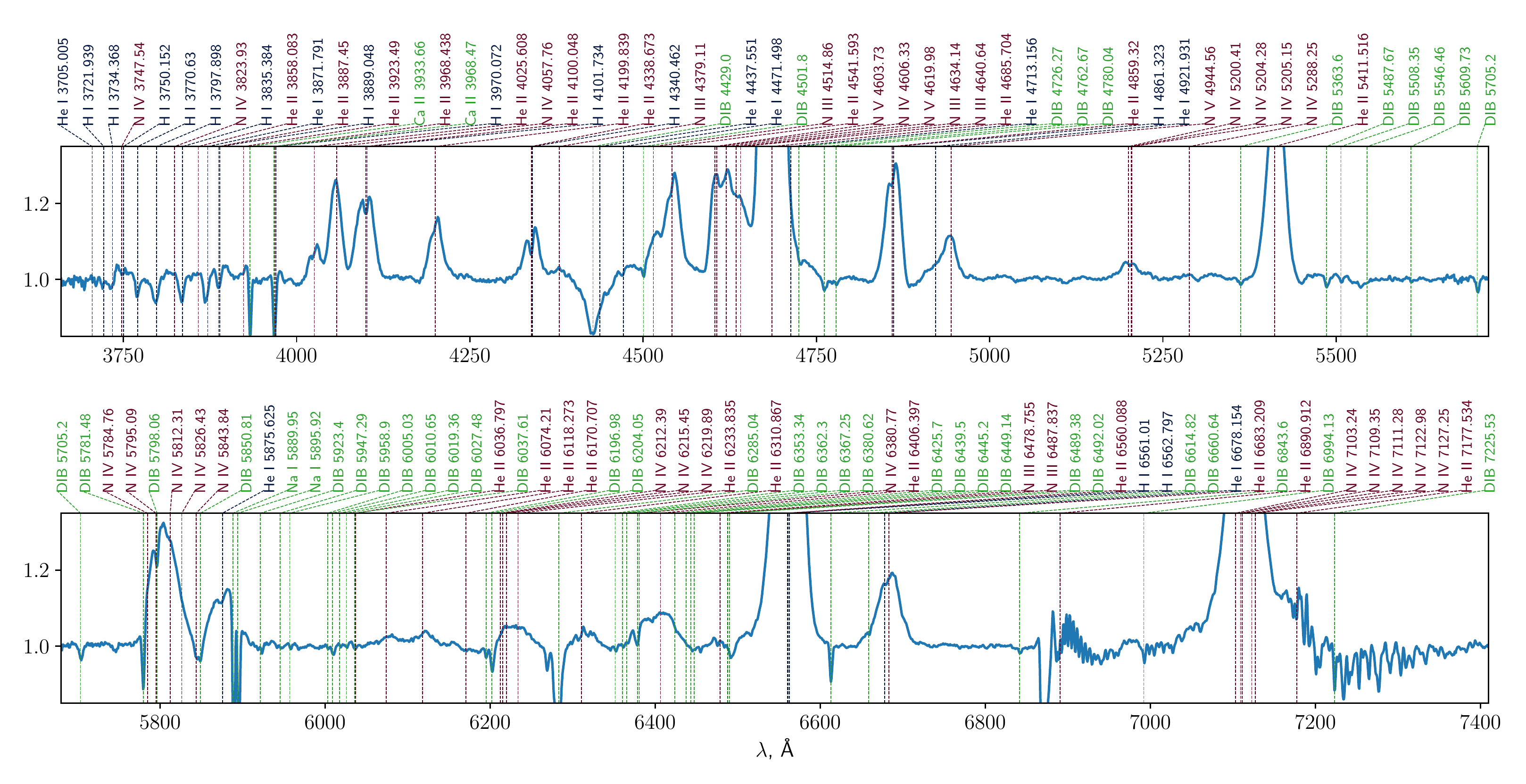}
   \caption{The averaged spectrum of CX Cep. The most prominent spectral details are marked. The interstellar lines are in green, absorption lines are in black and emission lines are in red.}
   \label{mean_CX}
\end{figure*}

Fig. \ref{mean_CX} shows the averaged (excluding radial velocity shifts) spectrum of CX Cep.
The strongest lines in the CX Cep spectrum belong to ionized helium ($\lambda$4686, $\lambda$4859, $\lambda$5411, $\lambda$6560, etc.). Also, noticeable are \ion{N}{iv} lines  ($\lambda$4058, blend $\lambda$7112), \ion{N}{v} ($\lambda$4603, $\lambda$4619, $\lambda$4944) and weak \ion{N}{iii} ($\lambda$4097, blends $\lambda$4511+4515, $\lambda$4634+4640) features. Comparing the relative line intensities of different nitrogen ions, the spectrum of the Wolf-Rayet star can be classified between WN4.5 and WN5, which is quite consistent with the WN5 class indicated for CQ Cep by other researchers (e.g. \citealp{Lewis1993}).

In contrast to CQ Cep, hydrogen Balmer absorption lines are seen in the CX Cep spectrum in counter-phase with WR emissions, which are reliably related to the secondary star. The H$\alpha$ - H$\delta$ lines are blended with strong \ion{He}{ii} emissions, which is the likely reason of their profiles variability and distortion measured in previous spectroscopic studies. The H$\epsilon$ absorption is superimposed with the Ca II H line formed in the interstellar medium and therefore does not show variations with the binary phase. The lines H$_9$-H$_{11}$ are comparatively 'clean' and deep enough for measurements of the O-star radial velocity.

The spectral class of the secondary component is estimated using the equivalent line width ratio $EW$($\lambda$4471 \ion{He}{i}) / $EW$($\lambda$4542 \ion{He}{ii}). To determine them, the average spectrum was calculated and reduced for the O-star velocity derived from H$_9$-H$_{11}$ lines. The lines were fitted with Gaussians on top of the background represented by the straight line between the assumed profile boundaries. The following equivalent widths were measured:
$EW$($\lambda$4471 \ion{He}{i}) = 0.102$\pm$0.017 \AA, $EW$($\lambda$4542 \ion{He}{ii}) = 0.319$\pm$0.003 \AA    
yielding $EW$($\lambda$4471 \ion{He}{i}) / $EW$($\lambda$4542 \ion{He}{ii}) = 0.32$\pm$0.05. This  is consistent with the value 0.33$\pm$0.07 found by \cite{Lewis1993} and allows us to attribute the star to O5 spectral class.

\begin{table}
    \centering
    \scriptsize
    \caption{Semi-amplitudes of radial velocities and $\gamma$-velocities for various spectral lines in the spectrum of CX Cep measured by different authors}
    \begin{tabularx}{\columnwidth}{XXcXXc}
        \hline
        mean MJD & Line & Type & $K$, km s$^{-1}$ & $\gamma$, km s$^{-1}$ & Ref.$^b$ \\
        \hline
        31748  & \ion{He}{ii}, $\lambda$4686 & Emis. & 290 &  & H48 \\
        31748  & \ion{He}{ii}, $\lambda$4686 & Emis. & 302 & 12 & B66 \\
        43879  & \ion{N}{iv}, $\lambda$4058  & Emis. & 300 & -93 & MC81 \\
          & \ion{He}{i}, \ion{H}{i}                 & Abs. & 130 & -93 &  \\
        47051  & H$_9$, $\lambda$3833    & Abs. &  $227\pm 25$  & $-21.1\pm0.4$  & L93 \\
          & H$_9$+H$_{10}$, $\lambda$3833     & Abs. &  $179\pm 19$  & $-29.8\pm0.4$  &  \\
          & \ion{He}{i}, $\lambda$3888       & Abs. &  $ 44\pm 18$  & -1492          &  \\ 
          & H$_8$,   $\lambda$3889       & Abs. &  $242\pm 16$  & $ -58.1\pm0.2$  &  \\ 
          & \ion{N}{iv}, $\lambda$4058       & Emis. & $306\pm 22$  & $-114.1\pm0.3$ &  \\
          & H$\delta$, $\lambda$4102  & Abs. &  $230\pm 12$  & $   -80\pm0.2$    &  \\
          & \ion{He}{ii}, $\lambda$4200      & Abs. &  $291\pm 23$  & $  10.1\pm0.2$   &  \\
          & \ion{He}{ii}, $\lambda$4200      & Emis. & $409\pm 14$  & $  16.3\pm0.2$   &  \\
          & H$\gamma$, $\lambda$4340  & Abs. &  $236\pm 12$  & $ -87.2\pm0.2$  &  \\
          & \ion{He}{ii}, $\lambda$4542      & Abs. &  $215\pm 14$  & $ -55.0\pm0.2$  &  \\
          & \ion{He}{ii}, $\lambda$4542      & Emis. & $405\pm 14$  & $  24.5\pm0.2$   &  \\
          & \ion{N}{v}, $\lambda$4603        & Emis. & $338\pm 10$  & $  -9.4\pm0.1$   &  \\
          & \ion{N}{v}, $\lambda$4619        & Emis. & $318\pm 10$  & $   7.5\pm0.1$   &  \\
          & N III, $\lambda$4642      & Emis. & $348\pm 15$  & $-101.6\pm0.1$ &  \\
          & \ion{He}{ii}, $\lambda$4686      & Emis. & $344\pm 10$  & $  50.0\pm0.1$   &  \\
          & \ion{He}{ii}, $\lambda$4859      & Emis. & $406\pm 13$  & $  44.1\pm0.1$   &  \\
          & H$\beta$, $\lambda$4861   & Abs.  &  $259\pm 11$  & $-132.2\pm0.2$ &  \\
    59496 & H$_9$-H$_{11}$            & Abs.  & $173 \pm 4$ & $-71 \pm9$ & $\dag$ \\
          & \ion{He}{ii}, $\lambda$4685      & Emis. & $243 \pm 3$ & $ 58 \pm3$ &    \\
          & \ion{He}{ii}, $\lambda$4859      & Emis. & $353 \pm 5$ & $ 65 \pm4$ &    \\
          & \ion{He}{ii}, $\lambda$5411      & Emis. & $323 \pm 5$ & $ 88 \pm3$ &    \\
          & \ion{He}{ii}, $\lambda$6560      & Emis. & $262 \pm 3$ & $ 43 \pm3$ &    \\
          & \ion{N}{iv}, $\lambda$4057       & Emis. & $317 \pm 4$ & $ -85\pm8$ &    \\
          & \ion{N}{iv}, $\lambda$7115       & Emis. & $322 \pm 3$ & $ -39\pm3$ &    \\
          & \ion{N}{v}, $\lambda$4603        & Emis. & $358 \pm 5$ & $ -51\pm3$ &    \\
          & \ion{N}{v}, $\lambda$4619        & Emis. & $326 \pm 3$ & $ -36\pm4$ &    \\
          & \ion{N}{v}, $\lambda$4944        & Emis. & $322 \pm 4$ & $-119\pm5$ &    \\
    \hline 
    \multicolumn{6}{c}{$^b$
    H48=\cite{Hiltner1948}, B66=\cite{Bracher1966}, MC81=\cite{Massey1981},} \\
    \multicolumn{6}{c}{
    L93=\cite{Lewis1993}, 
    $\dag$ = present paper}
    \end{tabularx}
    \label{vel_compare_CX}
\end{table}

Our measurements are in agreement with the archival data (Table \ref{vel_compare_CX}). The radial velocity semi-amplitude $K_{O}$ is derived from H$_9$-H$_{11}$ lines ($K_{O} = 173\pm 4~\mbox{km s}^{-1}$), and  $K_{WR}$ is the mean value derived from the \ion{N}{iv} and \ion{N}{v} lines (except for the \ion{N}{v} $\lambda$4603 line; $K_{WR} = 320\pm 5~\mbox{km s}^{-1}$). Assuming a circular orbit, we obtain $a_{WR}\cdot \sin{i} = 13.44 \pm 0.21~R_{\odot},~a_{O}\cdot \sin{i} = 7.27 \pm 0.17 ~R_{\odot};~
M_{WR}\cdot\sin^3 i = 9.3 \pm 0.5~M_{\odot},~M_{O}\cdot\sin^3 i= 17.2 \pm 0.4~M_{\odot},\quad q = \frac{M_{WR}}{M_O} = 0.54 \pm 0.03.$

Taking $i= 61^{\circ}$ from \cite{Hutton2009}, we find 
$a_{WR} = 15.4 \pm 0.5~R_{\odot},~a_{O} = 8.3 \pm 0.5~R_{\odot},~a = 23.7 \pm 1.0~R_{\odot};~M_{WR} = 13.9 \pm 0.7~M_{\odot},~M_{O} = 25.7 \pm 0.6~M_{\odot}.$

\section{Search for evolutionary orbital period changes for CQ Cep and CX Cep}

\subsection{CQ Cep}
The (O-C) diagram of CQ Cep 
can be interpreted by different models,
including the secular monotonic or non-monotonic period decrease, the rotation of the apse line and the presence of a third massive star in a long-period orbit. In the previous studies \citep{Kartasheva2006,Koenigsberger2017} the (O-C) diagram was calculated using the moments of eclipse minima from archival data. Despite the apparent abundance of data, the use of moments of minima only seems unreliable. A more reliable result can be obtained by a modified Hertzsprung method in which the light curves measured for different epochs are compared with a template light curve.
This method was first applied by \cite{Antokhina1981}. The CQ Cep period decrease was found to be $\dot{P} = - 0.019\pm 0.006~\mbox{s yr}^{-1}$, corresponding to a low WR mass-loss rate of $\dot{M}_{WR} = (0.69\pm 0.22)\cdot 10^{-5}~M_{\odot}$ yr$^{-1}$.
Later on \cite{Antokhina1987} used additional observations by \cite{Kreiner1983} and \cite{Kreiner1985} to determine the orbital period ($P = 1.641248^d$) and its decrease rate $\dot{P} = - 0.014\pm 0.004~\mbox{s yr}^{-1}$.

In the present paper, the (O-C) diagram is calculated using 
%photometric works in which the authors provided observation logs or average light curves, 
all available photometric data, the data from automated surveys ASAS and ASAS-SN \citep{Shappee2014,Kochanek_2017} and observations published by the American Association of Variable Star Observers (AAVSO) (see Table \ref{refs_CQ} for references).  The light curve obtained by \cite{Hiltner1950} was adopted as a template. This standard curve was matched with each mean light curve. 
%as a result of which there were positions in which the curves were as close to each other as possible. 
As a measure of proximity of the template and individual light curves, we used the criterion $\chi^2$ by calculating the primary minimum moment error at a significance level of 5\%.
Taking the model of the orbital period evolution with a constant rate $\dot{P}$, the difference between the observed primary minimum at a given epoch (O) and the calculated one (C) can be written as a quadratic equation $(O-C) = A\cdot E^2+B\cdot E+C$, where $\dot{P} = 2\cdot A$. The coefficients of the equation are found by the least squares method.

\begin{figure}
  \centering
  \includegraphics[scale=0.55]{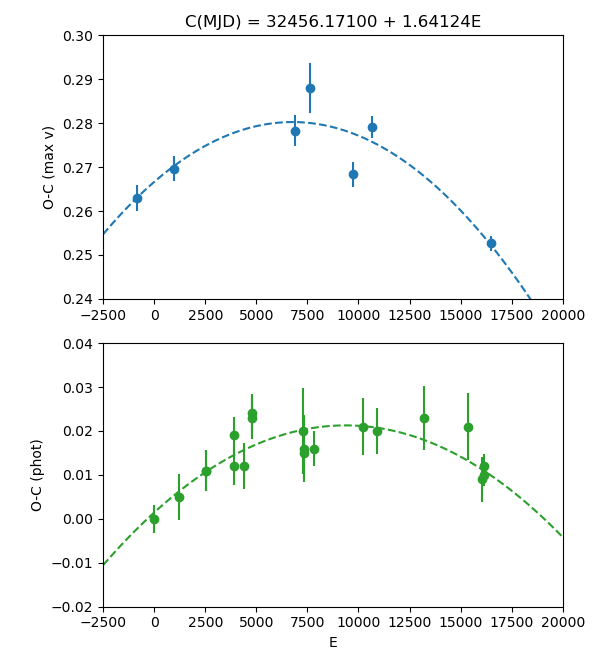}
  \caption{(O-C) diagrams for CQ Cep. The upper panel is the moments of the maximum radial velocity. The lower panel shows the phase shifts of the primary minimum relative to \protect\cite{Hiltner1950} observations.}
 \label{oc_cq1}
\end{figure}

The bottom panel in Fig. \ref{oc_cq1} shows the (O-C) diagram for primary minima %Despite some variation, the points are 
which is well fit by a parabola with a negative coefficient for the quadratic term suggesting the orbital period decrease over the time period covered by these observations. 

\begin{table}
    \centering
    \scriptsize
    \caption{Photometric observations of CQ Cep used to calculate the (O-C)-diagram}
    \begin{tabularx}{\columnwidth}{XXcXc}
    \hline
        mean MJD & (O-C) & Points & Type & Ref. \\
    \hline
        32451 & 0.0             & 382  & Phot., V   & \cite{Hiltner1950}     \\
        34415 & 0.005$\pm$0.005 & 176  & Phot., V   & \cite{Ishchenko1963}   \\
        36610 & 0.011$\pm$0.005 & 162  & PE, B      & \cite{Tchugaynov1960}  \\
        36610 & 0.011$\pm$0.005 & 162  & PE, V      & \cite{Tchugaynov1960}  \\
        38862 & 0.012$\pm$0.004 & 224  & PE, B      & \cite{Guseynzade1969}  \\
        38863 & 0.019$\pm$0.004 & 223  & PE, V      & \cite{Guseynzade1969}  \\
        39698 & 0.012$\pm$0.005 & 39   & PE, V      & \cite{Semeniuk1968}    \\
        40290 & 0.023$\pm$0.005 & 194  & PE, B      & \cite{Kartasheva1972}  \\
        40290 & 0.024$\pm$0.004 & 204  & PE, V      & \cite{Kartasheva1972}  \\
        44428 & 0.020$\pm$0.010 & 56   & PE, B      & \cite{Antokhina1981}   \\
        44522 & 0.016$\pm$0.008 & 82   & PE, V      & \cite{Antokhina1981}   \\
        44526 & 0.015$\pm$0.005 & 567  & PE, V      & \cite{Kreiner1983}     \\
        45306 & 0.016$\pm$0.004 & 398  & PE, V      & \cite{Kreiner1985}     \\
        49246 & 0.021$\pm$0.007 & 25   & Vis.       & AAVSO                  \\
        50326 & 0.020$\pm$0.005 & 40   & Phot., V   & AAVSO                  \\
        54165 & 0.023$\pm$0.001 & 49   & CCD, V     & AAVSO                  \\
        57670 & 0.021$\pm$0.008 & 115  & CCD, V     & ASAS-SN                \\
        58772 & 0.009$\pm$0.005 & 127  & CCD, V     & AAVSO                  \\
        58949 & 0.012$\pm$0.003 & 223  & CCD, V     & AAVSO                  \\
        58951 & 0.010$\pm$0.002 & 1155 & CCD, g     & ASAS-SN                \\
        \end{tabularx}
    \label{refs_CQ}
\end{table}

The orbital period change rate $\dot{P}$ can also be obtained by comparing the radial velocity curves in different epochs with each other. The upper panel in Fig. \ref{oc_cq1} shows the phases of the moments of the maximum radial velocity in the line \ion{N}{iv} $\lambda$4058 with the same ephemeris as the light curves used to calculate the photometric (O-C) diagram (given in the title of Fig. \ref{oc_cq1}). The zero epoch in this ephemeris corresponds to the moment of the primary minimum found by \cite{Hiltner1950}, the period is chosen to best-fit the data.

\begin{table}
    \centering
    \scriptsize
    \caption{Spectroscopic observations of CQ Cep used to calculate the (O-C)-diagram}
    \begin{tabularx}{\columnwidth}{XXXc}
    \hline
        mean MJD & (O-C) & Points & Ref. \\
    \hline
        31044  &  0.2648$\pm$0.0029  & 50  & \cite{Hiltner1944}     \\
        34012  &  0.2719$\pm$0.0029  & 58  & \cite{Bappu1977}       \\
        43770  &  0.2854$\pm$0.0035  & 26  & \cite{Leung1983}       \\
        44968  &  0.2945$\pm$0.0057  & 16  & \cite{Kartasheva1985}  \\
        48446  &  0.2760$\pm$0.0027  & 32  & \cite{Marchenko1995}   \\
        49939  &  0.2875$\pm$0.0025  & 56  & \cite{Harries1997}     \\
        59504  &  0.2645$\pm$0.0017  & 199 & Present paper               \\
    \end{tabularx}
    \label{oc_sp_CQ}
\end{table}

Table \ref{parab_CQ} lists the coefficients of the (O-C) parabolas and the inferred values of $\dot{P}$. It can be seen that the $\dot{P}$ found by both methods coincide within the errors allowing us to conclude that there is a secular decrease in the orbital period of CQ Cep.

\begin{table}
    \centering
    \scriptsize
    \caption{The parameters of the (O-C) quadratic fits in Fig.\ref{oc_cq1}}
    \begin{tabularx}{\columnwidth}{lXXlc}
    \hline
     & A & B & C & $\dot{P}$, s\;yr$^{-1}$ \\
    \hline
        O-C (vel.) & $(-3.0\pm0.8)\cdot10^{-10}$ & $(4.6\pm1.2)\cdot10^{-6}$ & $0.269\pm0.004$ & $-0.019\pm0.005$  \\
        \hline
        O-C (phot.) & $(-2.3\pm0.3)\cdot10^{-10}$ & $(4.2\pm0.6)\cdot10^{-6}$ & 
        $0.0014\pm0.0021$ & $-0.0145\pm0.0018$\\
    \end{tabularx}
    \label{parab_CQ}
\end{table}

\subsection{CX Cep}

\begin{figure}
  \centering
  \includegraphics[scale=0.55]{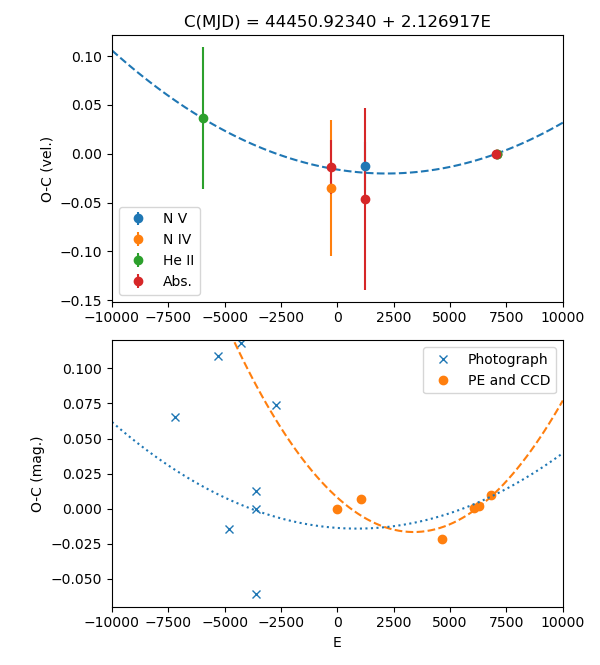}
  \caption{(O-C)-diagrams for CX Cep. The upper panel shows the moments of the maximum radial velocity in lines of different ions. The lower panel shows the primary minima phases.}
 \label{oc_cx1}
\end{figure}

For CX Cep there are much less photometric observations than for CQ Cep, and the heterogeneity of spectroscopic measurements carried out at different times does not enable a rigorous analysis. To calculate the (O-C) diagram, we used archival primary minima (see Table 7). %To compile this table, we used the tool "O-C gateway"\ \citep{}. 
Data includes times of minima found on archival photographic plates from 1898-1981 \citep{Kurochkin1985}, results of photoelectric measurements by \cite{Lipunova1981} and \cite{Stickland1988}, and minima times published by members of the Federal German Association for Variable Stars (BAV). To search for changes in the orbital period from spectral data differences were found between the moments of maximum radial velocity for each archived curve and the curve in the same line constructed from our spectral data (Table \ref{oc_sp_CX}). The results are shown in Fig. \ref{oc_cx1}. When estimating the rate of change of the orbital period from  photometric observations, the primary minima obtained by photoelectric or CCD photometry were considered separately (orange dots in the bottom panel of Fig. \ref{oc_cx1}) or together with the photographic minima with higher uncertainty from \cite{Kurochkin1985} (blue crosses).

The previous CX Cep spectral studies produced radial velocities in different spectral lines, therefore our (O-C) diagram was constructed by a different method. First, using the new data, the average radial velocity curves for each ion were constructed and superimposed on the archive curves. The relative shift of the curves and errors were calculated in the way as for the photometric (O-C) using the $\chi^2$ criterion (the significance level 5\%). A similar procedure was applied only to the radial velocity curves related to the same ion, because, firstly, the radial velocities in the lines of different ions shows a phase shift, and, secondly, the shapes of radial velocity curves for different ions differ from each other due to the effect of false eccentricity which can appear in emission lines of ions with a lower ionization potential.

Table \ref{parab_CX} lists the coefficients of the quadratic fits to the constructed (O-C) diagrams and $\dot{P}$ inferred from them. Among the three estimates presented, the one obtained from the maximum radial velocity phases appears to be the most reliable. Note that all estimates give a positive coefficient $A$ implying an increase in the orbital period of CX Cep.

\begin{table}
    \centering
    \scriptsize
    \caption{Photometric observations of CX Cep used to calculate the (O-C)-diagram}
    \begin{tabularx}{\columnwidth}{XXXc}
    \hline
        MJD & (O-C) & Type & Ref. \\
    \hline
        33217.2800	&  0.109   &  Phot., V	&  \cite{Kurochkin1985}  \\
        34229.4300	& -0.014   &  Phot., V	&  \cite{Kurochkin1985}  \\
        35348.4700	&  0.118   &  Phot., V	&  \cite{Kurochkin1985}  \\
        36796.5200	& -0.060   &  Phot., V	&  \cite{Kurochkin1985}  \\
        36807.3100	&  0.012   &  Phot., V	&  \cite{Kurochkin1985}  \\
        36809.4100	& -0.000   &  Phot., V	&  \cite{Kurochkin1985}  \\
        38698.2700	&  0.073   &  Phot., V	&  \cite{Kurochkin1985}  \\
        44096.3720	&  0.007   &  PE, V     &  \cite{Lipunova1981}   \\
        44451.4230	& -0.022   &  PE, V     &  \cite{Lipunova1981}   \\
        46714.4780	&  0.000   &  PE, V     &  \cite{Stickland1988}  \\
        54290.4946  &  0.002   &  CCD, V    &  BAV                   \\
        57338.4150  &  0.010   &  CCD, V    &  BAV                   \\
        58982.5416  &  0.011   &  CCD, V    &  BAV                   \\
    \end{tabularx}
    \label{refs_CX}
\end{table}

\begin{table}
    \centering
    \scriptsize
    \caption{Spectroscopic observations of CX Cep used to calculate the (O-C)-diagram}
    \begin{tabularx}{\columnwidth}{XlXXc}
    \hline
        MJD & (O-C) & Points & Ion & Ref. \\
        \hline
        31747 &  0.036$\pm$0.004  & 70  & \ion{He}{ii}      & \cite{Hiltner1948} \\                              
        43879 & -0.014$\pm$0.026  &  7  & \ion{H}{i} (abs.) & \cite{Massey1981}   \\
        43879 & -0.035$\pm$0.021  &  7  & \ion{N}{iv}       & \cite{Massey1981}   \\                        
        47050 & -0.014$\pm$0.005  & 57  & \ion{H}{i} (abs.) & \cite{Lewis1993}   \\
        47050 & -0.012$\pm$0.004  & 57  & \ion{N}{v}        & \cite{Lewis1993}   \\                            
        59445 &  0.0              & 77  & \ion{H}{i} (abs.) & Present paper         \\
        59464 &  0.0              & 134 & \ion{N}{v}        & Present paper         \\
        59475 &  0.0              & 182 & \ion{He}{ii}      & Present paper        \\
        59496 &  0.0              & 182 & \ion{N}{iv}       & Present paper         \\
    \end{tabularx}
    \label{oc_sp_CX}
\end{table}

\begin{table}
    \centering
    \scriptsize
    \caption{Parameters of the quadratic fits in Fig.\ref{oc_cx1}\\}
    \begin{tabularx}{\columnwidth}{lXXlc}
    \hline
     & A & B & C & $\dot{P}$, s\;yr$^{-1}$ \\
    \hline
        O-C (vel.) & $(8.51\pm0.15)\cdot10^{-10}$ & $(-3.7\pm0.6)\cdot10^{-6}$ & $(-1.6\pm0.6)\cdot10^{-2}$ & $0.054\pm0.009$  \\
        \hline
        O-C (phot., & \multirow{2}{*}{$(2.1\pm1.1)\cdot10^{-9}$} & \multirow{2}{*}{$(-1.5\pm0.8)\cdot10^{-5}$} & 
        \multirow{2}{*}{$(7.8\pm0.8)\cdot10^{-3}$} & \multirow{2}{*}{$0.13\pm0.07$}\\
         PE \& CCD) & & & \\
        \hline
        O-C (phot.,  & \multirow{2}{*}{$(6\pm5)\cdot10^{-10}$} & \multirow{2}{*}{$(-1\pm5)\cdot10^{-6}$} & 
        \multirow{2}{*}{$(-1\pm4)\cdot10^{-2}$} & \multirow{2}{*}{$0.04\pm0.03$}\\
        all) & & & \\
    \end{tabularx}
    \label{parab_CX}
\end{table}

\section{Discussion. Estimations of $\dot{M}$}

\subsection{CQ Cep} 

To estimate the rate of the orbital period decrease, we can take the intersection of the confidence intervals of two independent estimates obtained in the present paper (see Table \ref{parab_CQ}): $\dot{P} = -0.0151\pm 0.0013$ s\;yr$^{-1}$. The mass ratio $q = M_{WR}/M_{O}\approx 0.6$ is in contradiction with the previous studies 
The possible reason for the overestimation of $q$ in previous studies discussed above can be the attribution of \ion{He}{i} orbital shifts 
%\pk{removal - what is it? Unclear!} 
to the OB star despite large velocity shifts. Note that in \cite{Marchenko1995}, where careful attention was given to weak absorptions around $\lambda\approx 4100~\mbox{\AA}$, there are noticeably lower variations (in comparison with \ion{He}{i}) of the radial velocity semi-amplitude in \ion{Si}{iv} $\lambda$4116 line, the account of which gives $q$ close to our estimate. However, this fact, on the contrary, motivated the authors to remove this line from the analysis. \cite{Kartasheva1985} also changed their conclusion about the component masses.
Apparently, the above authors chose an overestimated $q$ under the influence of earlier studies in which the mass ratio was determined by questionable methods. In addition, given the compactness of the system, the proximity of the OB star radius to its Roche lobe and the complex gas dynamics of stellar winds, one should be careful when interpreting the CQ Cep radial velocity curves by assuming the point-like components.

\begin{figure}
   \centering
   \includegraphics[scale=0.45]{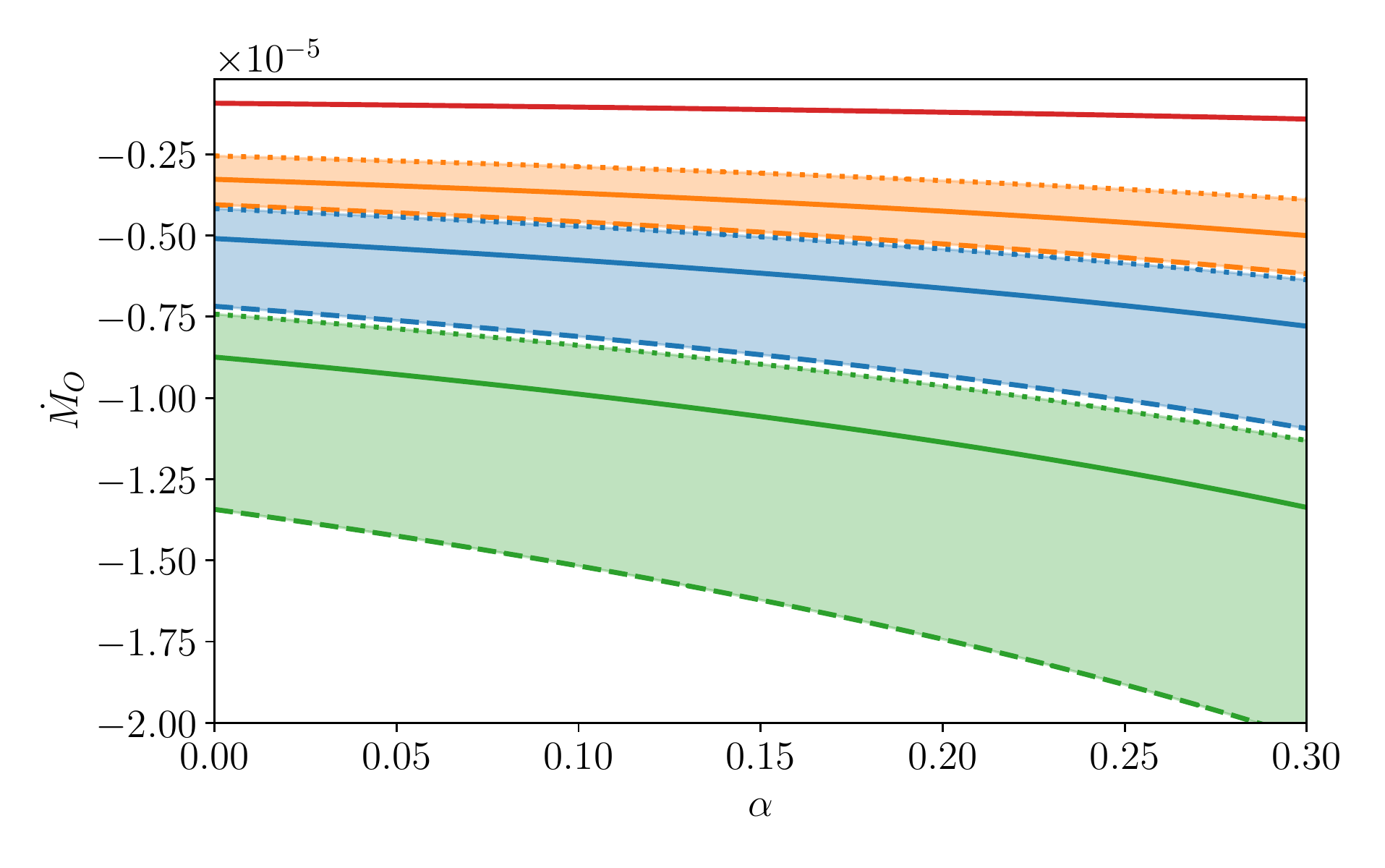}%
   \caption{The OB-star mass loss rate $\dot{M}_{OB}$ in CQ Cep for different  $q$ and $R_{OB},~R_{WR}$ as a function of the total mass-loss rate fraction $\alpha$ in the radial stellar wind.\\
   \emph{Red line:} critical $\dot{M}_{OB}$ for $\dot{M}_{WR,w} = 0$,\\
   \emph{Orange area:} $\dot{M}_{WR,w} = -5\cdot 10^{-6}~M_{\odot}~\mbox{yr}^{-1}$,\\
   \emph{Blue area:} $\dot{M}_{WR,w} = -1\cdot 10^{-5}~M_{\odot}~\mbox{yr}^{-1}$,\\
   \emph{Green area:} $\dot{M}_{WR,w} = -2\cdot 10^{-5}~M_{\odot}~\mbox{yr}^{-1}$;\\
   \emph{Solid lines} -  $R_{OB} = 8.0~R_{\odot}$, $q = 0.7$, $R_{WR} = 6.8~R_{\odot}$ (from \protect\cite{Antokhina1988});\\
   \emph{Dashed lines} - $R_{OB} = 7.6~R_{\odot}$, $q = 0.6$ (from the present paper), $R_{WR} = 0$, \\
   \emph{Dotted lines} - $R_{OB} = 7.6~R_{\odot}$, $q = 0.6$ (from the present paper), $R_{WR} = 6.0~R_{\odot}$ (according to \protect\cite{Eggleton1983}).}
   \label{CQ_Mdot}
\end{figure}

\begin{table*}
    \centering
    %\scriptsize
    \caption{Results of $\dot{M}_O$ ($M_{\odot}\mbox{yr}^{-1}$) calculations for CX Cep ($\dot{M}_{WR} = x \cdot \dot{M}_O$).}
    \begin{tabularx}{\textwidth}{XXccc}
    \hline
    Masses                 & Radii                                      & $x = 10$ & $x = 100$ & $x = 1000$ \\
    \hline
    $M_O = 25.7 M_{\odot}$, $M_{WR} = 13.9 M_{\odot}$    & $R_O = 8.5 R_{\odot}$, $R_{WR} = 7.0 R_{\odot}$    & $-8.60\cdot10^{-7}$ & $-9.15\cdot10^{-8}$ & $-9.21\cdot10^{-9}$ \\ \cline{2-5} 
    (This paper)  %
    & $R_O = R_{WR} = 0$  & $-5.25\cdot10^{-7}$ & $-5.72\cdot10^{-8}$ & $-5.77\cdot10^{-9}$ \\
    \hline
    $M_O = 36.8 M_{\odot}$, $M_{WR} = 26.4 M_{\odot}$    & $R_O = 8.5 R_{\odot}$, $R_{WR} = 7.0 R_{\odot}$    & $-1.15\cdot10^{-6}$ & $-1.24\cdot10^{-7}$ & $-1.25\cdot10^{-8}$ \\ \cline{2-5} 
    \citep{Hutton2009} %
    & $R_O = R_{WR} = 0$  & $-8.38\cdot10^{-7}$ & $-9.13\cdot10^{-8}$ & $-9.21\cdot10^{-9}$ \\
    \end{tabularx}
    \label{mdot_CX}
\end{table*}

Using the found values of the component masses and $\dot{P}$ of the CQ Cep system, we can estimate the mass exchange and stellar wind rates. To explain the shortening of the orbital period of CQ Cep, we consider two competing processes driving the orbital evolution: the mass flow from a more massive component to a less massive one decreasing $P$ (in the case of conservative mass exchange), and fully non-conservative radial matter outflow from the system increasing the orbital separation. 
This model was analyzed  in papers  \cite{Antokhina1981,Antokhina1987}. According to \cite{Antokhina1981}, the change in the orbital period CQ Cep can be represented as:
\beq{pdot_simple}
\frac{\dot{P}}{P} = -3\frac{\dot{M}_{WR}^{(2)}}{M_{WR}} (q-1) + 3\frac{\dot{M}_{O}^{(2)}}{M_{WR}} (q-1) + 2 \frac{\dot{M}_{WR}^{(1)}}{M_{WR}} \frac{q}{1+q},
\eeq
where index (1) denotes the mass loss due to the radial matter outflow (assuming  $\dot{M}_{WR}^{(1)} \gg\dot{M}_{O}^{(1)}$), and index (2) marks the mass exchange rate between the components.
Note that this formula was used to analyze the observed value of $\dot{P}$ 
by assuming the mass ratio from \cite{Niemela1980} ($q = M_{WR}/M_{OB} = 1.21$), 
implying the mass outflow from the WR to OB star.
Since in this paper we came to the conclusion that $q < 1$, we assume the opposite direction of mass transfer. Using the derived $\dot{P}$ and assuming $q = 0.6$, $\dot{M}_{WR}^{(2)} = 0$, we can estimate the minimum value of $\dot{M}_{O}^{(2)}$, as well as constraints on the $\dot{M}_{WR}^{(1)}$ to $\dot{M}_{O}^{(2)}$ ratio:
$$ |\dot{M}_{O}^{(2)}| \geq (0.97\pm 0.09)\cdot10^{-6}~M_{\odot}\mbox{yr}^{-1},~|\dot{M}_{O}^{(2)}|\geq 0.62|\dot{M}_{WR}^{(1)}|.$$

These estimates seem unconvincing because the expected mass loss rate $\dot{M}$ from stars of these spectral classes is by an order of magnitude higher. In addition, our analysis (see Appendix) demonstrates that when taking into account the finite sizes of the components, a situation is possible where $\dot{P} < 0$ without mass exchange between the components (i.e., due to stellar winds only). Therefore, it is necessary to recalculate $\dot{M}$ taking into account the stellar wind from components with finite radii. We can use  $R_{OB} = 8.0~R_{\odot},~R_{WR}= 6.8~R_{\odot}$ obtained by \cite{Antokhina1988} from the modelling of the CQ Cep light curves for  $q = 0.7$. Additionally, the OB-star size can be taken as the equivalent Roche lobe radius \citep{Eggleton1983}. In this analysis, there are free parameters $\alpha$ and $x$ characterizing the fraction of the radial OB star mass loss in the total mass loss of the OB star, and the ratio $\dot{M}_{WR,w}/\dot{M}_{OB}$, respectively (see Appendix). We believe that the fraction of the radial stellar wind outflow from the OB star does not exceed 20-30\% of the total mass loss rate $\dot{M}_{OB}$ ($\alpha\leq 0.2-0.3$), while $\dot{M}_{WR}$ is significantly higher compared to $\dot{M}_{OB}$. Fig. \ref{CQ_Mdot} shows the range of possible $\dot{M}_{OB}$ for different WR mass loss rates  $\dot{M}_{WR}$ ($0,~-5\cdot10^{-6},~-1\cdot10^{-5},~-2\cdot10^{-5}~M_{\odot}~\mbox{yr}^{-1}$), fixed sizes of stars and the ratio $\dot{P}/P$ found in this paper. The most realistic is the case with $\dot{M}_{WR} =10^{-5}~M_{\odot}~\mbox{yr}^{-1}$ with the parameter $\alpha\sim 0.1$. In this case (with realistic assumptions about the radii of stars) $\dot{M}_{OB}$ turns out to be about $5\cdot10^{-6}~M_{\odot}~\mbox{yr}^{-1}$.

\subsection{CX Cep}

The spectral analysis of the CX Cep system enabled us to construct reliable radial velocity curves and to derive orbital parameters and component masses. The obtained parameters are in agreement with previous studies. The new result is the estimate of the orbital period increase rate $\dot{P} =0.054\pm0.009~\mbox{s\, yr}^{-1}$ derived by comparing the radial velocity curves.

%Due to the relatively simple behavior of the spectra, in which 
The CX Cep spectra show no traces of direct interaction of the components, and it can be assumed that the CX Cep orbit evolves in a pure Jeans mode. Using the $\dot{P}$ and $M_{WR}$ found in this paper and assuming that the matter outflow from the system is entirely due to the powerful WR wind, its mass loss rate is 
$\dot{M}_{WR} = -\frac{1}{2} \frac{\dot{P}}{P} (M_{WR}+M_O) = -(5.8\pm 0.4)\cdot 10^{-6}~M_{\odot}\mbox{yr}^{-1}.$

To improve this estimate, we take into account the O-star stellar wind mass loss  and the finite sizes of the components. %As an estimate of the radii of stars,
We assume $R_O = 8.5~R_{\odot}$, $R_{WR} = 7.0~R_{\odot}$ calculated by \cite{Hutton2009} based on BVRI photometry. Note, however, that there is a significant difference in the masses obtained here and in \cite{Hutton2009}, so we used different stellar masses. The results are presented in Table. \ref{mdot_CX} for three values $x=10,100,1000$ (notation and formulas see Appendix A) and two assumptions about the star radii (point-like stars and spherical stars with radii according to \citealp{Hutton2009}). The calculations show that taking into account these effects slightly increases  $\dot{M}_{WR}$ up to a value of the order of $-1\cdot 10^{-5}M_{\odot}~\mbox{yr}^{-1}$, which is in good agreement with the model values $\dot{M}_{WR}$ for WR stars of similar spectral classes.

\begin{table}
    \centering
    \scriptsize
    \caption{Observed and derived parameters of CQ Cep and CX Cep}
    \label{t:final}
    \begin{tabularx}{\columnwidth}{lXlX}
\hline
                          & CQ Cep                   &                  & CX Cep                \\
\hline                                               
$P$, days                 & $1.641239$    & $P$, days        & $2.127066$   \\
                          & $\pm000051$ &                  & $\pm000377$ \\
$K_{WN6}$, km s$^{-1}$           & $297\pm9$            & $K_{WN5}$, km s$^{-1}$  & $320\pm5$             \\
$f_{WN6}(M)$, $M_{\odot}$ & $4.46\pm0.27$            & $K_{O5}$, km s$^{-1}$   & $173\pm4$             \\
$q = M_{WN6}/M_{O9-B0}$      & $\approx 0.6$           & $M_{WN5}\cdot \sin^3{i}$, $M_{\odot}$ & $9.3\pm0.5$    \\
$a_{WN6}\cdot \sin{i}$, $R_{\odot}$  & $9.6\pm0.3$ & $M_{O5} \cdot \sin^3{i}$, $M_{\odot}$ & $17.2\pm0.4$   \\
                          &                          & $a_{WN5}\cdot \sin{i}$, $R_{\odot}$   & $13.44\pm0.21$ \\  
                          &                          & $a_{O5} \cdot \sin{i}$, $R_{\odot}$   & $7.27\pm0.17$  \\      
\hline                                                                   
                          & $i = 60^{\circ}~(q=0.6)$   &                        & $i = 61^{\circ}$      \\
$M_{WN6}$, $M_{\odot}$    & $10.8\pm0.6$                & $M_{WN5}$, $M_{\odot}$ & $13.9\pm0.7$          \\
$M_{O9-B0} $, $M_{\odot}$    & $18.0\pm1.0$                & $M_{O5} $, $M_{\odot}$ & $25.7\pm0.6$          \\
$a_{WN6}$, $R_{\odot}$    & $11.1\pm0.4$              & $a_{WN5}$, $R_{\odot}$ & $15.4\pm0.5$          \\
$a_{O9-B0} $, $R_{\odot}$    & $6.7\pm0.7$                 & $a_{O5} $, $R_{\odot}$ & $8.3\pm0.5$           \\
\hline
$\dot{P}$, s\;yr$^{-1}$           & $-0.0151\pm0.0013$   & $\dot{P}$, s\;yr$^{-1}$  &  $+0.054\pm0.009$  \\

    \end{tabularx}
\end{table}
\section{Conclusion}

In the present paper, we report on new spectroscopic observations of two short-period eclipsing WR+OB eclipsing binary systems  CQ Cep and CX Cep obtained in 2020-2023 with the TDS spectrograph on the 2.5-m telescope of the Caucasian Mountain Observatory of SAI MSU and describe the data processing and interpretation. The main goal of our study has been to search for a secular evolutionary change in the orbital period in these systems using the available spectral data. We found traces of the absorption lines in the CQ Cep spectra which allowed us to estimate the binary mass ratio  $q=M_{WR}/M_O \approx 0.6$, the component masses and the orbital size. Here new photometric observations are required to improve the obtained estimates of the binary parameters.
For CX Cep, the radial velocities of both components have been constructed enabling the 
determination of component masses and orbital size (see Table \ref{t:final}).

Using the radial velocity curves of CQ Cep and CX Cep obtained in the present paper and radial velocity measurements at earlier epochs from the literature, we derived the orbital period change rates in these binaries $\dot P$ which turned out to be in agreement with estimates from light curve analysis (see Table \ref{t:final}). Our finding validates the use of the spectroscopic method for the $\dot P$ estimates in WR+OB binaries. This method can be used to estimate $\dot P$ in a wide class of spectral non-eclipsing binary WR+OB systems. 

We have also developed a new method of the dynamical assessment of the mass-loss rate from the components of WR+OB systems from the observed values of $\dot P$ with taking into account finite size of the stars. Using this method, we give improved estimates of the mass-loss rates from WR stars $\dot M_{WR}$ in CQ Cep and CX Cep about $\sim 10^{-5}~M_\odot$ yr$^{-1}$. 

The application of the spectroscopic method can provide  dynamical estimates of the mass-loss rate for several dozen WR stars with different masses and spectral classes. If the number of epochs of spectral observations of non-eclipsing binary WR+OB systems is insufficient for the $\dot P$ and $\dot M$ estimates, constraints on $\dot P$ and $\dot M$ can be derived. In these cases, new spectral observations will provide additional epoch for future determination of $\dot P$ and $\dot M$ from spectroscopic data. 
Therefore, further spectral studies of spectroscopic binary WR+OB systems are of great interest.

\section*{Acknowledgements}

The authors thank the anonymous referee for useful notes. 
ISh, ACh and AD acknowledge the support from RSF grant 23-12-00092 (observations, data analysis).
The work of KP is partially supported by the International Space Science Institute (ISSI) in Bern, through ISSI International Team project 512 Multiwavelength View on Massive Stars in the Era of Multimessenger Astronomy.
The research is supported by the Scientific and Educational School of M.V. Lomonosov Moscow State University 'Fundamental and Applied Space Research'. The Program of Development of Moscow University is also acknowledged.

\section*{Data Availability} 

Data of spectroscopic observations are available on reasonable request from the authors.

%\bibliographystyle{mnras}
%\bibliography{refs}

\begin{appendix}
\label{app}
\section{Orbital period change due to stellar wind mass loss from spherical synchronously rotating components with finite radii}

Consider two spherical stars with masses $M_1$, $M_2=q M_1$ ($q$ is the mass ratio) and radii\footnote{Here we assume the 'radius' of a star as the distance from star's barycenter at which the stellar wind is launched.} $R_1$, $R_2$ in a circular orbit with separation $a$. The orbital period is $P$. We assume that the primary component $M_1$ loses mass via spherically symmetric stellar wind at the rate $\dot M_{1,w}$. If the radius $R_1$ of the primary component is reasonably close to the Roche lobe, there also can be mass transfer onto the secondary component through the inner Lagrangian point at the rate $\dot M_{1,t}$. The secondary component accretes the mass transferred from the primary component and can also lose mass via spherically symmetric stellar wind at the rate $\dot M_{2,w}=x\dot M_{1,w}$. Assuming no mass loss via the outer Lagrangian points, the mass balance reads:
\beq{a:mass1}
\dot M_1=\dot M_{1,w}+\dot M_{1,t}=\alpha \dot M_1+\beta \dot M_1,\quad \alpha+\beta=1\,,
\eeq
\beq{a:mass2}
\dot M_2=\dot M_{2,w}-\dot M_{1,t}=x \dot M_1-\beta \dot M_1\,.
\eeq

The orbital angular momentum is 
\beq{a:J}
J=\frac{M_1M_2}{M}\omega a^2=\frac{M_1M_2}{M}\sqrt{GMa}, \quad M=M_1+M_2\,.
\eeq
Here the orbital frequency $\omega = 2\piup/P$ is related to the total mass $M$ and orbital separation $a$ via the third Kepler's law $\omega^2=\frac{GM}{a^3}$.
The angular momentum balance reads:
\beq{a:dJ}
\frac{\dot J}{J}=\frac{\dot M_1}{M_1}+\frac{\dot M_2}{M_2}-\frac{\dot M}{M}+\frac{1}{2}\frac{\dot a}{a}=
\frac{\dot J_\mathrm{out}}{J}
\eeq
where $\dot J_\mathrm{out}$ is the angular momentum loss rate from the system due to the stellar wind. In terms of fractional orbital period change, using the third Kepler's law, we can write:
\beq{a:dP}
\frac{\dot P}{P}=-\frac{1}{2}\frac{\dot M}{M}+\frac{3}{2}\frac{\dot a}{a}=\frac{\dot M}{M}-3\frac{\dot M_1}{M_1}-3\frac{\dot M_2}{M_2}+3\frac{\dot J_\mathrm{out}}{J}\,.
\eeq

We need to specify the angular momentum loss term $\dot J_\mathrm{out}$. Considering stars as point-like masses, the angular momentum carried away by spherically symmetric wind would be $\dot J_i=\dot M_i\omega a_i^2$, $a_i=a(M_{3-i}/M)$, for the $i$-th component ($i=1,2$). However, if radii of the components are taken into account, in the synchronously rotating case (which is appropriate in a close binary system) we should integrate specific angular momentum from each area element over the entire surface, $\dot J_\mathrm{out,i}=
\int dS (d \dot M_i/dS)\omega \rho_i^2$. Here $\rho_i$ is the distance from the element to the binary orbital rotation axis passing through the system barycenter. Let us use the spherical coordinates from stars' centers ($r, \theta,\phi$) such that $dS=R_i^2\cos\theta d\phi d\theta$. Note that $\rho^2(\theta,\phi)=a_i^2+R_i^2\cos\theta^2-2a_iR_i\cos\theta\cos\phi$ by the cosine theorem. After integrating over the entire surface, we obtain 
\beq{a:dJfull}
\frac{d J_{\mathrm{out},i}}{dt}=\dot M_{i,w}\omega \left(a_i^2+\frac{2}{3}R_i^2\right), \quad i=1,2.
\eeq
Clearly, the second term is important when the stellar radii are not small compared to the distance to the system barycenter.

Using \eqn{a:mass1}-\eqn{a:dJfull} and making algebraic rearrangements, we finally arrive at the equation relating the fractional orbital period change $\dot P/P$ to the full mass loss from the primary $\dot M_1/M_1$ in the form:
\beqa{a:dPP}
&\displaystyle\frac{\dot P}{P}=-\frac{\dot M_1}{M_1}\left\{3+3\frac{x-\beta}{q}-\frac{\alpha+x}{1+q}\right.\\
&\displaystyle\left.-3\frac{1+q}{q}\left(
\alpha\left[\myfrac{q}{1+q}^2+\frac{2}{3}\myfrac{R_1}{a}^2\right]+
x\left[\myfrac{1}{1+q}^2+\frac{2}{3}\myfrac{R_2}{a}^2\right]
\right)
\right\}\nonumber
\eeqa
Terms in the second line of \Eq{a:dPP} are responsible for the angular momentum loss via spherically symmetric stellar wind from synchronously rotating components. Setting zero $\alpha=0, x=0$ (no stellar wind mass loss) and $\beta=1$ reduces \Eq{a:dPP} to the standard expression for conservative mass transfer: $\dot P/P=-3(\dot M_1/M_1)(1-1/q)$. 
Setting $\beta=0$, 
$R_1=R_2=0$ and $\alpha=1$ reduces \Eq{a:dPP} to the standard expression for spherically symmetric wind mass loss from point-like stars (the Jeans mode): $\dot P/P=-2\dot M/M$.

Without mass exchange ($\beta=0$, $\alpha=1$), the fractional orbital period change due to stellar wind mass loss from non-zero spherical synchronously rotating binary components reads:
\beq{a:dPP0}
\frac{\dot P}{P}=-2\frac{\dot M_1}{M_1}\frac{1+x}{1+q}
\left\{1-\frac{(1+q)^2}{q(1+x)}\left[\myfrac{R_1}{a}^2+
x\myfrac{R_2}{a}^2\right]
\right\}.
\eeq
The second term in the curved brackets vanishes for point-like masses. For stars with finite radii, for large and small $q$ it can exceed unity, leading to \textit{negative} $\dot P$. This is in sharp contrast with the standard treatment of spherically symmetric wind mass loss via the Jeans mode from point-like stars (always positive $\dot P$). For example, in the limiting case $x\ll 1$ (insignificant mass loss from the secondary), $\dot P>0$ for $R_1/a<1/2$ for any $q$. If $R_1/a>1/2$, there are critical mass ratios below and above which the period derivative is negative, $q_{1,2}^{(cr)}=-1+(1/2)(a/R_1)^2\pm (a/R_1)^2\sqrt{(1/4)(a/R_1)^2-1}$.

If the size of the primary component is close to the Roche lobe, $R_1=a f(q)$, where  $f(q)$ is the standard function of the mass ratio (for example, using Eggleton's approximation $\displaystyle f(q)=\frac{0.49 q^{2/3}}{0.6 q^{2/3}+\ln(1+q^{1/3})}$). Then \Eq{a:dPP0} turns into
\beq{a:dPP1}
\frac{\dot P}{P}=-2\frac{\dot M_1}{M_1}\frac{1+x}{1+q}
\left\{1-\frac{(1+q)^2}{q(1+x)}f^2(q)\left[1+
x\myfrac{R_2}{R_1}^2\right]
\right\}.
\eeq

\end{appendix}
\label{lastpage}
\end{document}